\documentclass[preprint,12pt]{elsarticle}

\usepackage{amsmath}
\usepackage{amssymb}
\usepackage{graphicx}
\usepackage{xcolor}
\usepackage{booktabs}
\usepackage{array}
\usepackage{multirow}
\usepackage{ragged2e}
\usepackage{url}
\usepackage[hidelinks]{hyperref}
\hypersetup{
  breaklinks = true,
  bookmarksnumbered = true,
  pdftitle   = {Temporal generalization and explanation stability of control flow
                graph neural networks for malware detection},
  pdfauthor  = {Md. Asif Sajeed and Md. Nazrul Islam Mondal and
                Md. Ashraful Hossen Akash},
  pdfkeywords= {malware detection, graph neural networks, control flow graphs,
                temporal generalization, concept drift, explainability}
}

\usepackage[a4paper,left=2.5cm,right=2.5cm,top=2.5cm,bottom=2.5cm]{geometry}

\graphicspath{{./}}

\newcolumntype{R}[1]{>{\RaggedRight\arraybackslash}p{#1}}

\usepackage{etoolbox}
\AtBeginEnvironment{tabular}{\footnotesize}

\newcommand{\eqtag}[1]{{\setlength{\fboxsep}{1.7pt}%
  \colorbox{black!75}{\textcolor{white}{\scriptsize\textbf{#1}}}}}

\journal{Journal of Information Security and Applications}

\begin{document}

\begin{frontmatter}

\title{Temporal generalization and explanation stability of control flow
graph neural networks for malware detection}

\author[ruet]{Md. Asif Sajeed\corref{cor1}}
\ead{asifsajeed01@gmail.com}
\cortext[cor1]{Corresponding author.}

\author[ruet]{Md. Nazrul Islam Mondal}
\ead{nimbd109@gmail.com}

\author[ruet]{Md Ashraful Hossen Akash}
\ead{ashraful.hossen.akash@gmail.com}

\affiliation[ruet]{organization={Department of Computer Science and Engineering,
                                 Rajshahi University of Engineering \& Technology},
                   city={Rajshahi},
                   postcode={6204},
                   country={Bangladesh}}

\begin{abstract}
Malware detection is a critical task in cybersecurity, and graph neural networks over
control flow graphs have shown promising results for it. However, detectors are usually
evaluated on a random split of a corpus collected over a single period, which cannot
show how well a model generalizes to later samples. This study addresses that
limitation with a strict temporal split: every model is trained on one period and
scored once on a later one. Two corpora of control flow graphs, each node carrying 37
features, were extracted statically from 1,989 Windows portable executables: 459
graphs from 2024--2025 for training and 223 from 2026 for evaluation. Twelve variants
and a flat-feature control were trained on the earlier corpus. The choice of
message-passing operator changes robustness to the shift significantly, and every
pairwise gap that survives correction separates an aggregating architecture from one
built around a learned attentional readout. The ranking also reverses: the flat
control, which sees node features but no topology, is the best in-distribution model
and among the worst across the boundary, so a conventional benchmark would have
rejected message passing. Neither recalibration nor ensembling substitutes for the
operator choice. Attributions do not shift, but explanation validity is
architecture-specific, and the most accurate operator on the later corpus is the
hardest to explain. An architecture derived from the finding matches the best searched
operator without search. The shift affects both malware and benign classes alike, so
these are results about robustness to distribution shift, not malware evolution.
\end{abstract}


\end{frontmatter}

\section{Introduction}
\label{sec:intro}

\subsection{Motivation and problem}
\label{sec:motivation}

AV-TEST records on the order of hundreds of thousands of newly seen malicious programs
per day \cite{avtest2026}. No team of analysts can triage that volume, so detection has
moved from hand-written signatures to statistical models, and signature matching in any
case fails by construction against polymorphic and packed samples, which produce a fresh
byte pattern per victim \cite{ugartepedrero2015sok}.

Among the representations a learned detector can be built on, the \emph{control flow
graph} (CFG) is currently the most active. A Windows Portable Executable can be
disassembled, without ever being run, into a graph whose nodes are basic blocks of
machine instructions and whose edges are the jumps and calls between them, and a graph
neural network (GNN) can be trained to classify that graph \cite{yan2019dgcnn,
ling2022malgraph, peng2024malgne, shokouhinejad2025stacked}. Unlike a flat feature
vector, which counts entropies, imported libraries and header fields but discards which
piece of code runs after which, a CFG preserves the structure of the program, and the
reported numbers are strong.

The difficulty lies in how those numbers are produced. Almost every published
CFG-based GNN detector is evaluated on a random split of a single corpus, which
measures whether a model can recognize the malware it was built from, not whether it
still works next year on programs written after training finished. Security machine
learning has named this failure \emph{temporal bias}: TESSERACT
\cite{pendlebury2019tesseract} shows that violating time-ordering constraints
inflates reported performance, and Arp et al.\ \cite{arp2022dosdonts} catalog
temporal snooping as one of the field's most widespread pitfalls, in a line of
criticism running back to Sommer and Paxson \cite{sommer2010outside}. The CFG--GNN
literature has largely not adopted the remedy.

\begin{figure}[!t]
  \centering
  \includegraphics[width=1.0\textwidth]{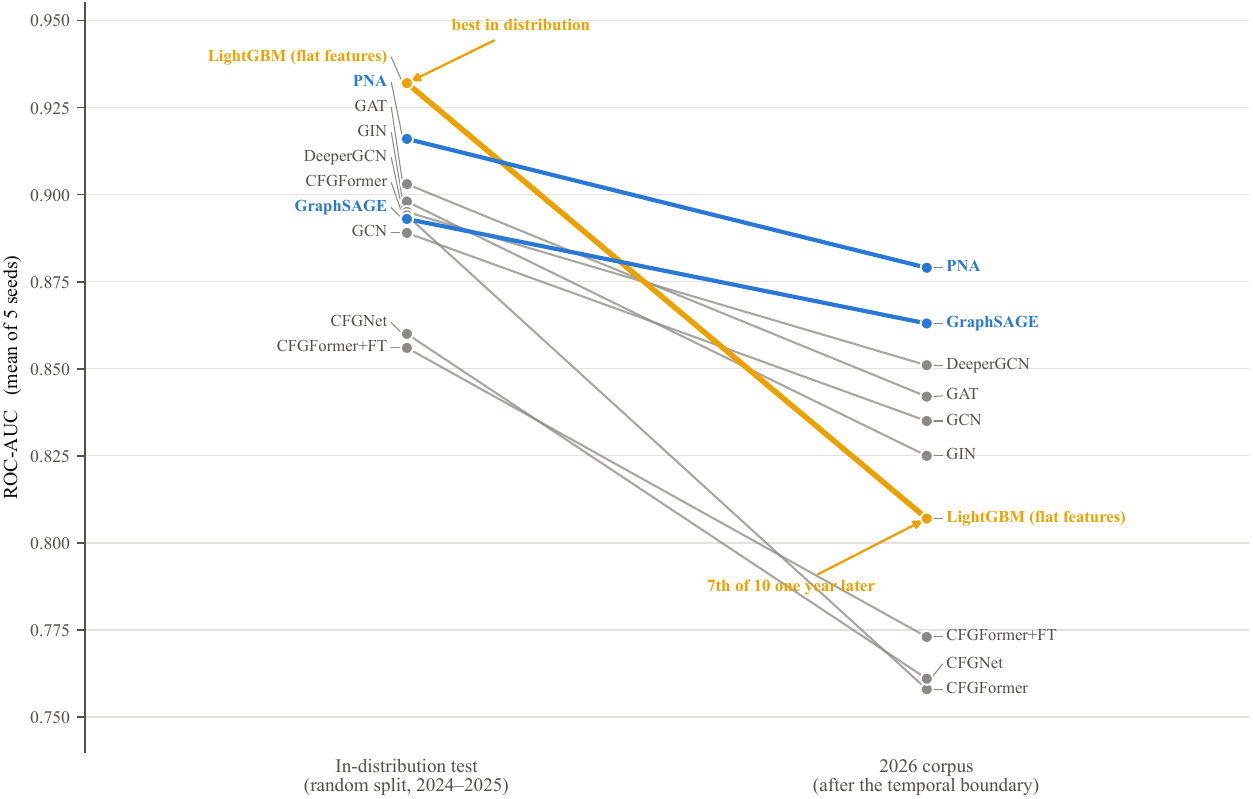}
  \caption{Rank reversal between the in-distribution and 2026 evaluations. The
           flat-feature control (no topology) is the strongest model in
           distribution and among the weakest across the temporal boundary.}
  \label{fig:reversal}
\end{figure}

This matters more than a correction of over-optimistic figures would suggest, because
the bias is not uniform across designs. If a random split inflated every architecture
equally, a leader-board built on one would still rank designs correctly and the numbers
could simply be discounted. Figure~\ref{fig:reversal} shows that this is not what
happens. We trained ten models (nine GNN variants and a gradient-boosted control given
graph-level summary statistics, that is everything the GNNs see \emph{except} the
topology) on binaries collected in 2024--2025, and scored them on a held-out slice of
that corpus and on binaries collected in 2026. In distribution the control is the best
of the ten, at AUC 0.932; across the boundary it falls to 0.807, while the best GNN
holds 0.879.

A benchmark run the conventional way would have concluded that message passing was not
worth its cost, and a temporal split reverses that conclusion. The claim concerns the
evaluation protocol rather than any one architecture, which is why we treat a temporal
split as an \emph{instrument} rather than a stricter grading scheme: it carries
information a random split cannot express.

\subsection{Research gap and questions}
\label{sec:gap-questions}

Two literatures bear on that observation and do not meet. The graph-learning literature
compares message-passing operators, but in distribution \cite{kipf2017gcn,
velickovic2018gat, xu2019gin, hamilton2017graphsage, corso2020pna}. The security
machine-learning literature establishes what a time-respecting evaluation requires
\cite{pendlebury2019tesseract, arp2022dosdonts}, but for feature-vector detectors, and
does not compare architectures under it. Within CFG-based detection, temporal evaluation
appears once, in MalGraph \cite{ling2022malgraph}, which does compare operators across a
date boundary, but on a malware-only split at near-saturated accuracy and without
measuring degradation, concluding that the operator does not matter, a finding we take
up in Section~\ref{sec:d-prior}. The closest explainability work evaluates explanation
stability under synthetic perturbation on a single corpus
\cite{shokouhinejad2025consistency}. Section~\ref{sec:rw-gap} develops the position.
Three questions follow.

\begin{enumerate}
  \item[\textbf{RQ1}] Does the choice of message-passing operator change how
  well a detector survives the passage of time? Published work compares GNN
  architectures on in-distribution accuracy; whether the most accurate operator is
  also the most durable has not been measured.
  \item[\textbf{RQ2}] When a model degrades over time, does its explanation
  degrade with it? Explainability work on malware GNNs evaluates explanations on a
  single corpus under synthetic perturbation \cite{shokouhinejad2025consistency}, not
  under real deployment drift. Whether an attribution still means the same thing a
  year later is unknown.
  \item[\textbf{RQ3}] If the answer to RQ1 is a ranking, can that ranking be
  designed from? An architecture specified \emph{from} the finding, rather than
  selected off a leader-board, tests whether the finding carries design information.
\end{enumerate}

Answering all three required two corpora of Windows binaries either side of a hard date
boundary, a static CFG extractor, twelve GNN variants and a flat control trained under a
protocol in which the later corpus is scored once per run, a controlled factorial over
the two factors RQ1 implicates, a reproduction of the closest published system, and an
attribution study on both sides of the boundary.

A random split ranks a topology-free control
above every graph model; a temporal split reverses that ranking; the reversal is not
uniform but tracks how each architecture builds its graph vector; a controlled factorial
isolates how much of that belongs to the readout; an attribution study asks whether
explanations move with accuracy; and an architecture specified from the resulting
finding is built and tested against the field. Sections~\ref{sec:r-shift} to
\ref{sec:r-explain} follow that order.

\subsection{Contributions}
\label{sec:contributions}

\begin{enumerate}
  \item The message-passing operator changes drift robustness, and the effect
  is statistically real. Omnibus tests on temporal degradation are significant in
  two independent executions with separate hyper-parameter searches, and seven
  pairwise gaps on out-of-distribution AUC survive multiplicity correction. Every one
  separates an architecture that builds its graph vector by parametric aggregation
  from one built around a learned attentional readout. The one prior operator
  comparison across a date boundary \cite{ling2022malgraph} reaches the opposite
  conclusion, under conditions Section~\ref{sec:d-prior} sets out
  (Section~\ref{sec:r-arch}).

  \item Graph structure only pays for itself out of distribution. The flat
  control is the best in-distribution model in the study and among the worst across
  the boundary (Figure~\ref{fig:reversal}); the two best GNNs beat it on the later
  corpus by $+0.072$ and $+0.055$ AUC. This is the methodological core of the paper
  and is independent of which GNN wins (Section~\ref{sec:r-flat}).

  \item Attributions survive the boundary, but explanation validity is a
  property of the architecture, not of the data. No architecture-by-concept
  comparison shifts significantly across the boundary, nor do the topology controls;
  yet the two architectures we studied explain themselves incompatibly. One
  grounds in no behavioural concept at all; the other grounds strongly but in the
  opposite direction to the intuitive hypothesis, selecting small, quiet blocks rather
  than busy API-heavy ones; and the two agree at chance level. The operator that scores
  highest on the later corpus is the one hardest to explain locally
  (Section~\ref{sec:r-explain}).

  \item The drift finding is prescriptive. CFGNet-v2 (a PNA backbone with
  residual connections, Jumping Knowledge and a plain mean readout, every component
  chosen because a measurement pointed at it) matches the best searched operator in
  a twelve-variant field (a tie under a pre-registered noise floor, not a win) and
  beats five comparators under correction, including the flat control that CFGNet and
  both CFGFormer variants lost to. The two architectures we designed on intuition
  finished last and second to last; the one we designed on evidence finished first
  (Section~\ref{sec:r-derived}).

  \item Two supporting results from independent methods. A faithful
  reproduction of the closest published system \cite{shokouhinejad2025stacked} shows
  that its stacking gain does not reproduce under this protocol, and that ensembling
  three drift-fragile operators is significantly beaten by selecting one drift-robust
  operator. A controlled learning-curve study shows that for the two highest-capacity
  models, \emph{more} training data improves the in-distribution score while degrading
  the out-of-distribution one (Sections~\ref{sec:r-repro} and \ref{sec:r-lc}).
\end{enumerate}

\subsection{Scope and limitations}
\label{sec:scope}

Four boundaries qualify everything that follows. Each is revisited where it bites and
collected again in Section~\ref{sec:d-threats}.

The shift was measured per class, with the
benign class as a control, under a verdict rule fixed before the measurement. Benign
software drifts at least as much as malware (Section~\ref{sec:r-drift}), so a toolchain,
compiler or collection artifact explains it at least as well as malware evolution does.
Every result here is therefore about robustness to a measured distribution shift, never
about malware evolution. The boundary is clean and the corpora genuinely distinct, so
the question remains well posed; only its interpretation narrows.

The corpus is small and purpose-built: 1,989
binaries yield 682 usable graphs, because no public corpus supplies what this study
requires, namely raw Windows PE binaries, of both classes, carrying per-sample dates
(Section~\ref{sec:why-custom}). It is comparable in size to the CFG explainability work
this paper is positioned against, and our reproduction lands within two to four accuracy
points of the original authors' figures.

Every interval is a standard deviation over five seeds on
one fixed split, and therefore measures optimisation variance; Section~\ref{sec:stats}
defines the paired bootstrap we use alongside it for test-set sampling uncertainty.

Adversarial robustness is not evaluated here. An
attacker perturbing a binary and a population drifting over time are different
phenomena, and a detector can be robust to one and fragile to the other
(Section~\ref{sec:rw-adv}).

\subsection{Organisation}
\label{sec:organisation}

Section~\ref{sec:related} reviews related work and states the gaps.
Section~\ref{sec:problem} formalises what we measure and states the research questions
as hypotheses. Section~\ref{sec:method} describes the corpora, extraction, model space
and protocols. Section~\ref{sec:results} reports results, Section~\ref{sec:discussion}
discusses them and states threats to validity, and Section~\ref{sec:conclusion}
concludes.

\section{Related work}
\label{sec:related}

We review the four literatures this paper sits between (static detection and program
representations, CFG-based GNN detection, message-passing operators, and temporal
evaluation) together with the explainability work it extends, and close by stating the
gaps that produce the three research questions. We discuss works for what their
evaluations cover, since accuracies on different corpora and label sets are not
comparable. One note on terminology: several works describe angr- or IDA-recovered
graphs as \emph{dynamically constructed}, meaning recovered by an analysis engine rather
than declared in advance, not that the binary is executed. Every graph here is produced
without running anything.

\subsection{Static malware detection and control flow graphs}
\label{sec:rw-repr}

Signature matching is exact and fast but useless against anything unseen;
semantics-aware matching \cite{christodorescu2005semantics} replaced literal patterns
with behavioural templates at the cost of a human author per template, and machine
learning replaced template-writing with statistical generalisation. Broad surveys of
that transition \cite{ozkanokay2024survey} treat detection as flat classification
throughout, addressing neither graph structure nor interpretability.

Static flat-feature models dominate the large-scale end. EMBER \cite{anderson2018ember}
standardised the approach (header, section, import and byte-histogram statistics over
1.1 million PE files with a gradient-boosted tree) and remains the reference static
baseline; MalConv \cite{raff2018malconv} instead feeds raw bytes to a convolutional
network, and malware-as-image methods \cite{nataraj2011malwareimages} sit between the
two. All three describe \emph{what is in} a file rather than \emph{how it executes}, and
all degrade on packed samples. Our flat-feature control (Section~\ref{sec:flat}) is a
deliberate stand-in for this family, which is why the comparison in
Section~\ref{sec:r-flat} is between representations and not only between models.

Dynamic and behavioural methods run samples in an instrumented environment.
Behaviour-graph approaches \cite{kolbitsch2009effective} build dependency graphs over
system calls, but the more common modern treatment flattens execution into an API call
sequence and applies a sequence model, over corpora such as Mal-API-2019
\cite{catak2020malapi}, with convolutional and self-attentional recurrent detectors
\cite{lin2024maldetect} or Transformers over byte-pair-encoded call sequences
\cite{youssef2025transformer}. A flat sequence discards which function invokes which,
and neither line evaluates temporal generalisation.

The public corpora shaping this literature matter for a second reason: they determine
what an independent group can evaluate. BIG 2015 \cite{ronen2018big2015}, BODMAS
\cite{yang2021bodmas}, notable for carrying timestamps and family labels, and SOREL-20M
\cite{harang2020sorel} are the standard references, and vendor family labels are noisy
enough that normalisation tooling exists to reconcile them \cite{sebastian2016avclass}.
Section~\ref{sec:why-custom} explains why none supports the evaluation we require.

\subsection{CFG-based GNN malware detection}
\label{sec:rw-cfg}

Graph representations retain the structure the other families discard. Mitra et al.\
\cite{mitra2023survey} trace the line from string-encoded CFG blocks to learned graph
topology and close on extraction, encoding and explainability as the open problems,
without covering temporal generalisation. Shokouhinejad et al.\
\cite{shokouhinejad2025survey} cover the graph-learning and explainability side and
reach two conclusions that motivate this paper: explanation quality is evaluated
inconsistently, and temporal robustness essentially not at all. Yan et al.\
\cite{yan2019dgcnn} were among the first to apply a deep graph convolutional network
with a sorting-based readout to malware CFGs.

MalGraph \cite{ling2022malgraph} is the most architecturally complete system
in this space and the most important comparison point for RQ1. It represents an
executable as a hierarchical graph (a function call graph capturing inter-function
semantics, with a per-function control flow graph underneath) and learns over both
levels end to end with GraphSAGE, which is why GraphSAGE is in our candidate set. Of
the CFG-based GNN detectors we surveyed it is the only one whose evaluation includes
a time-based split, and it goes further than that suggests: it trains on malware
collected before 6 May 2020, compares four detectors spanning three representation
families under that split, and ablates its own backbone over GraphSAGE, GCN and
TransConv, concluding that the architecture ``is not sensitive to the choice of GNN
variants'' at 99.49, 99.93 and 99.97\% AUC. That conclusion is the opposite of ours,
and Section~\ref{sec:d-prior} addresses it directly. Three features of the design
bound how far it reaches: the split is time-based for \emph{malware only} (goodware
is split randomly at matching proportions), so the benign half of any shift is
removed by construction; all models score above 98\% AUC, leaving little room to
separate operators; and no in-distribution comparison is reported, so no degradation
is measured. Its node features are instruction counts, its extraction is static and
therefore evaded by packing, and it provides no explainability.

MalGNE \cite{peng2024malgne} attacks the node-encoding problem instead,
defining a rule-based instruction encoding that removes the out-of-vocabulary
behaviour of dictionary-based encodings such as Asm2Vec \cite{ding2019asm2vec} and
compressing it into a low-dimensional space; it retains 95.49\% accuracy at 16 node
dimensions for roughly 73\% less training time than 128. That out-of-vocabulary
concern is what motivates our fixed, vocabulary-free node encoding
(Section~\ref{sec:nodefeat}). MalGNE offers neither explainability nor temporal
analysis.

\subsection{Message-passing operators and readouts}
\label{sec:rw-ops}

The operators we compare come from the general graph-learning literature: GCN
\cite{kipf2017gcn}, degree-normalised neighbourhood smoothing; GraphSAGE
\cite{hamilton2017graphsage}, which keeps separate weights for a node's own state and
its neighbourhood mean and is inductive by construction; GAT \cite{velickovic2018gat},
which learns a weight per edge; GIN \cite{xu2019gin}, sum aggregation with a learnable
self-weight and provable 1-WL expressiveness; PNA \cite{corso2020pna}, which combines
four aggregators with three degree-based scalers so that degree enters the operator as
an input rather than something it must re-derive; DeeperGCN \cite{li2020deepergcn},
learnable softmax aggregation inside deep residual stacks. Our graph-transformer
variant follows GraphGPS \cite{rampasek2022gps}, which pairs local message passing
with a global attention channel. Supporting components include Jumping Knowledge
\cite{xu2018jk}, the gated attentional readout of gated graph sequence networks
\cite{li2016ggnn}, masked-label
message passing \cite{shi2021masked} as the global update in our graph-transformer
variant, and GraphCL \cite{you2020graphcl} for contrastive pretraining.

This body of work compares operators in distribution only, on benchmarks whose splits
are random by construction. The property it establishes (expressiveness, the ability to
distinguish graphs in the limit) differs from generalising from a few hundred training
graphs to next year's binaries. Which operator survives a distribution shift is a
question it does not set out to answer, and that gap produces RQ1.

\subsection{Temporal evaluation and concept drift}
\label{sec:rw-temporal}

Outside malware graphs, temporal bias is well established. Sommer and Paxson
\cite{sommer2010outside} argued early that security is unusually hostile to standard
evaluation practice, and Gama et al.\ \cite{gama2014drift} give the taxonomy of drift on
which Section~\ref{sec:shift} builds. TESSERACT \cite{pendlebury2019tesseract} makes the
requirements concrete (training must precede testing in time, class ratios must be
realistic, no tuning may read the test period) and shows that violating them inflates
reported performance substantially; our evaluation lock (Section~\ref{sec:lock})
implements them. Arp et al. \cite{arp2022dosdonts} survey thirty top-tier security
papers and find sampling bias and temporal snooping widespread. On the mitigation side,
Transcend \cite{jordaney2017transcend} uses conformal evaluation to detect when
predictions have become unreliable, and Chen et al.\ \cite{chen2023continuous} study
continuous retraining under drift for Android malware.

The gap is therefore not that temporal evaluation is unknown, but that it has not been
used to measure \emph{how much each design loses}. TESSERACT establishes the protocol
without comparing architectures; MalGraph compares them, but across a boundary that
moves only the malicious class, at accuracies too high to separate operators, and with
no in-distribution figure to difference against, so it reports which model scores best
after the boundary, not which degrades least across it. No work asks whether the
architecture choice interacts with the size of that loss (Section~\ref{sec:r-arch}), or
whether the resulting ranking can be designed from (Section~\ref{sec:r-derived}).

\subsection{Explainability}
\label{sec:rw-xai}

General-purpose attribution methods (LIME \cite{ribeiro2016lime}, SHAP
\cite{lundberg2017shap}, integrated gradients \cite{sundararajan2017ig}) do not transfer
directly to graphs, whose inputs are variable-sized and lack a fixed feature ordering.
GNNExplainer \cite{ying2019gnnexplainer} learns a soft node and edge mask preserving the
prediction on one instance; PGExplainer \cite{luo2020pgexplainer} amortises that across
instances. Two cautions are why we evaluate attribution against an explicit degree-based
null model rather than by inspection: attribution maps can look convincing while being
largely insensitive to the model that produced them \cite{adebayo2018sanity}, and
interpretability claims need measurable criteria rather than visual plausibility
\cite{doshivelez2017rigorous}.

In the malware setting, CFGExplainer \cite{herath2022cfgexplainer} was the
first to explain GNN classifications of control flow graphs, ranking node importance
through a surrogate trained on node embeddings. Shokouhinejad et al.\
\cite{shokouhinejad2025consistency} build node features from a rule-based
439-dimensional instruction encoding reduced by an autoencoder, compare GNNExplainer,
PGExplainer and three Captum strategies, and introduce an aggregation method fusing
the top explainers and a greedy edge-wise subgraph procedure. Their evaluation uses
accuracy, fidelity and \emph{consistency}, but is entirely in distribution on a single
corpus with no time-based split, so explanation stability under real drift is not
addressed and explanations are never linked back to known malicious behaviour. Those
two omissions are what RQ2 targets.

The closest published system is the attention-guided stacking ensemble of Shokouhinejad
et al.\ \cite{shokouhinejad2025stacked}, which we reproduce in
Section~\ref{sec:r-repro}. It stacks GCN, GIN and GAT base learners through an attention
meta-learner that both classifies and weights each base model's contribution, and fuses
their edge-level explanations using those weights. On its own 831-sample corpus it
reports 0.8614 accuracy against 0.8313, 0.8253 and 0.8373 for the base learners and
0.8293 for a plain average ensemble, at 0.939 AUC. It names the absence of any temporal
evaluation as a limitation, which makes it the natural test of whether \emph{ensembling
several operators} substitutes for \emph{choosing one robust operator}.

\subsection{Research gaps and positioning}
\label{sec:rw-gap}

Table~\ref{tab:rw-compare} positions these works by \emph{what was evaluated} rather
than by reported accuracy. Its three evaluation columns correspond to the three research
questions, and no row but the last carries all three: MalGraph has the temporal split
and an architecture comparison but no explainability, and the closest explainability
work has the architecture comparison and the explanations but no temporal split.

\begin{table}[!t]
  \centering
  \caption{The closest related work, compared by what its evaluation covers rather
           than by reported accuracy. \emph{Time} means the evaluation includes a
           time-ordered split; \emph{Arch.} means several model architectures are
           compared under that evaluation; \emph{XAI} means explanations are
           produced and evaluated.}
  \setlength{\tabcolsep}{4pt}
  \begin{tabular}{R{3.1cm}R{1.9cm}R{1.5cm}cccR{2.7cm}}
  \toprule
  \textbf{Work} & \textbf{Repre-} \textbf{sentation} & \textbf{Scale} &
  \textbf{Time} & \textbf{Arch.} & \textbf{XAI} &
  \textbf{Gap left open} \\
  \midrule
  Catak et al.\ \cite{catak2020malapi} & API sequence & 7.1k & No & No & No &
    No structure; families only \\
  Lin et al.\ \cite{lin2024maldetect} & API sequence & --- & No & No & No &
    Drift, evasion, imbalance \\
  Anderson and Roth \cite{anderson2018ember} & Flat PE & 1.1M & Partial & No & No &
    No structure; packing \\
  Ling et al.\ \cite{ling2022malgraph} & FCG + CFG & Large & Yes &
    Yes (3) & No & Malware-only split; no $\Delta$; no XAI \\
  Peng et al.\ \cite{peng2024malgne} & CFG & Large & No & Yes & No &
    No XAI; no drift \\
  Herath et al.\ \cite{herath2022cfgexplainer} & CFG & Moderate & No & No &
    Yes & No drift; no operator study \\
  Shokouhinejad et al.\ \cite{shokouhinejad2025consistency} & CFG & 831 & No &
    Partial & Yes & Consistency is in-distribution \\
  Shokouhinejad et al.\ \cite{shokouhinejad2025stacked} & CFG & 831 & No &
    Yes (3) & Yes & No drift; edge-level only \\
  Pendlebury et al.\ \cite{pendlebury2019tesseract} & Android features & 129k &
    Yes & No & No & Not graphs; no operator study \\
  \midrule
  \textbf{This paper} & CFG, 37-d nodes & 682 graphs &
    Yes & Yes (12) & Yes & --- \\
  \bottomrule
  \end{tabular}
  \label{tab:rw-compare}
\end{table}

\paragraph{Gap 1: the loss each operator incurs across a boundary is unmeasured} The
graph-learning literature compares operators in distribution; the security-ML literature
establishes temporal protocols without comparing operators; the one operator comparison
under a time-based split reports absolute accuracy rather than degradation, on a
malware-only boundary \cite{ling2022malgraph}; and the rest of the CFG--GNN literature
names drift as future work without measuring it \cite{lin2024maldetect, mitra2023survey,
youssef2025transformer, shokouhinejad2025consistency, shokouhinejad2025stacked}.
\emph{This becomes RQ1.}

\paragraph{Gap 2: explanation stability is evaluated in distribution only} It is never
evaluated across time and never grounded in behaviour. Where it is measured at all it is
measured on one corpus \cite{shokouhinejad2025consistency}, and whether two equally
accurate architectures agree on \emph{what} they attribute is not asked. \emph{This
becomes RQ2.}

\paragraph{Gap 3: architecture comparisons stop at a ranking} No work tests whether a
ranking carries enough information to specify a new architecture, which is the
difference between a descriptive result and a design rule. \emph{This becomes RQ3.}

\paragraph{Gaps this paper does not close} Every CFG--GNN work reviewed, including ours,
is binary rather than family-level, and every static pipeline, including ours,
under-represents packed binaries. We also do not evaluate adversarial
robustness\label{sec:rw-adv}: MalAOI \cite{peng2025malaoi} rewrites PE files by
inserting benign opcode sequences selected by reinforcement learning, preserving
maliciousness while evading CFG-based GNN detectors on BODMAS and SOREL-20M, and finds
that adversarial training reduces but does not remove the vulnerability. That threat
model is orthogonal to ours, but it is a serious weakness of exactly the
detectors we study.

\section{Problem formulation and evaluation framework}
\label{sec:problem}

This section fixes the objects and quantities we measure, so that
Section~\ref{sec:method} can describe how we obtained them without re-arguing what they
mean.

\subsection{Attributed control flow graphs}
\label{sec:acfg}

Disassembly turns the bytes of a Portable Executable's code sections into machine
instructions, grouped into \emph{basic blocks}, maximal straight-line sequences with a
single entry and exit. The \emph{control flow graph} is the directed graph
\begin{equation}
G = (V, E), \qquad V = \{\text{basic blocks}\}, \qquad E \subseteq V \times V
\label{eq:cfg}
\end{equation}
where $(u,v) \in E$ means control can pass from $u$ to $v$ by fall-through, jump or
call. Each node carries a feature vector $\mathbf{x}_i \in \mathbb{R}^{d}$ describing
the content of that block, so the object handed to a model is the attributed graph
$(\mathbf{X}, \mathbf{E}, y)$ with $\mathbf{X} \in \mathbb{R}^{|V| \times d}$ and a
binary label $y$. Detection is the graph classification problem of predicting $y$ from
$(\mathbf{X}, \mathbf{E})$.

Recovering a CFG statically is undecidable in general: indirect jumps, computed call
targets, self-modifying code and packed sections all defeat a static disassembler, and a
packed binary is a small unpacking stub around an encrypted blob, so static analysis
recovers only the stub \cite{ugartepedrero2015sok}. This is the direct cause of the
extraction losses in Section~\ref{sec:r-corpus}, of the node-count floor, and of a
survivor bias that qualifies every result here.

Almost every modern GNN instantiates the message-passing framework
\cite{gilmer2017mpnn}: each node repeatedly summarises its neighbours and folds that
into its own state, so after $L$ rounds its representation describes an $L$-hop
neighbourhood of code. Writing $\mathbf{h}_i^{(l)}$ for node $i$ after $l$ rounds, with
$\mathbf{h}_i^{(0)} = \mathbf{x}_i$,
\begin{equation}
\mathbf{m}_i^{(l)} = \bigoplus_{j \in \mathcal{N}(i)}
   \phi^{(l)}\!\left(\mathbf{h}_i^{(l)}, \mathbf{h}_j^{(l)}\right),
\qquad
\mathbf{h}_i^{(l+1)} = \psi^{(l)}\!\left(\mathbf{h}_i^{(l)}, \mathbf{m}_i^{(l)}\right)
\label{eq:mp}
\end{equation}
where $\phi$ builds a message, $\bigoplus$ is a permutation-invariant aggregator and
$\psi$ updates the node. Classification needs one vector per graph, so a \emph{readout}
collapses the node set,
\begin{equation}
\mathbf{h}_G = \mathrm{READOUT}\left(\{\mathbf{h}_i^{(L)} : i \in V\}\right),
\label{eq:readout}
\end{equation}
which a classification head maps to a prediction. Because a graph has no canonical node
ordering, both $\bigoplus$ and $\mathrm{READOUT}$ must be set functions.

These two choices are what we vary, and they matter more here than in many graph domains
because the graphs are large: at a median of roughly 5,900 nodes a mean-pooled readout
averages any locally distinctive signal away, which makes a learned attentional readout
the obvious response, and its behaviour under shift worth measuring rather than
assuming. A network of this form is at most as expressive as the one-dimensional
Weisfeiler--Leman test, reaching that ceiling only when $\bigoplus$ and $\psi$ are
injective \cite{xu2019gin}. Expressiveness in that sense is the ability to distinguish
graphs in the limit, a different property from generalising from a few hundred training
graphs to next year's binaries.

\subsection{Temporal generalisation and \texorpdfstring{$\Delta$AUC}{Delta AUC}}
\label{sec:temporal-gen}

A detector is trained once and deployed forward in time, so what matters is performance
on samples that did not exist when training finished. We make that operational with two
disjoint corpora separated by a hard date boundary: an earlier corpus
$\mathcal{D}_{\text{tr}}$, from which we draw the training, validation and
in-distribution test splits, and a strictly later $\mathcal{D}_{\text{te}}$ standing in
for deployment.

We then read two numbers for every model: $\mathrm{AUC}_{\text{ID}}$, on the held-out
slice of the earlier corpus, which is what a conventional random-split study would
report; and $\mathrm{AUC}_{2026}$, on the later corpus. \emph{Temporal degradation} is
their difference,
\begin{equation}
\Delta\mathrm{AUC} = \mathrm{AUC}_{\text{ID}} - \mathrm{AUC}_{2026}
\label{eq:delta}
\end{equation}
reported as a mean $\pm$ standard deviation over five seeds. We use two readings of
Eq.~\eqref{eq:delta} throughout and they should not be conflated: $\mathrm{AUC}_{2026}$
answers \emph{which model to deploy}, while $\Delta\mathrm{AUC}$ answers \emph{which
design degrades least}. A model can lead on one and not the other, and
Section~\ref{sec:r-derived} reports exactly that case.

\paragraph{The evaluation lock}
\label{sec:lock}
$\mathrm{AUC}_{2026}$ is only an estimate of deployment behaviour if nothing in the
model's construction has seen the later corpus. We therefore adopt one rule, which
every component obeys and which implements the time-ordering requirement of TESSERACT \cite{pendlebury2019tesseract}:

\begin{quote}
Every choice (architecture selection, hyper-parameters, early stopping,
learning-rate scheduling, decision threshold, feature standardisation) reads data
from before the boundary only. The later corpus is scored exactly once per run,
after the checkpoint is frozen.
\end{quote}

We assert the lock in code rather than leaving it as an intention;
Section~\ref{sec:protocol} gives the assertions. We state what it permits and forbids
once here and do not repeat it at each experiment.

\subsection{Characterising the distribution shift}
\label{sec:shift}

Degradation across a boundary is attributable to the boundary only if the boundary is a
genuine shift, and to malware evolution only if the shift is specific to malware. We
assume neither. The standard decomposition \cite{gama2014drift} separates
\emph{covariate shift}, $P_{\text{tr}}(X) \neq P_{\text{te}}(X)$ with the labelling rule
unchanged, from \emph{concept drift}, $P_{\text{tr}}(Y \mid X) \neq P_{\text{te}}(Y \mid
X)$. In malware both occur, and separating them empirically needs family labels our
corpora do not carry (Section~\ref{sec:data-limits}).

What is measurable without them is whether the shift is \emph{specific to the malicious
class}, so we measure it per class with benign software as a control. We reduce each
graph to 40 graph-level features (three size and density statistics plus the mean of
each of the 37 node features) and run two measurements per class.

The first asks whether individual features moved. A two-sample Kolmogorov--Smirnov test
\cite{massey1951ks} gives a statistic $D$; since the classes have unequal sample sizes
and raw $D$ is not comparable between them, we report $D/D_{\text{crit}}$ with
\begin{equation}
D_{\text{crit}} = 1.358\sqrt{\frac{n_1+n_2}{n_1 n_2}}
\label{eq:dcrit}
\end{equation}
at $\alpha=0.05$. We call a feature \emph{notable} if $D>0.15$, a threshold we fixed
before the measurement, and correct $p$-values for multiplicity with the
Benjamini--Hochberg procedure \cite{benjamini1995fdr}.

The second asks the stronger question of whether the two corpora are separable at all,
jointly: we train a \emph{domain classifier} (a random forest of 300 trees with balanced
class weights, 5-fold cross-validated) to predict which corpus a graph came from. An AUC
near 0.5 means the corpora are indistinguishable within that class.

The verdict rule was fixed in advance: if the benign control drifts as much as or more
than malware, the shift cannot be attributed to malware evolution and must be reported
as \texttt{CONFOUNDED}. Section~\ref{sec:r-drift} reports what the rule returned, and it
scopes every subsequent claim. Because the rule was stated before the measurement, the
resulting narrowing of scope is a finding rather than a concession.

\subsection{Research questions and hypotheses}
\label{sec:hypotheses}

We restate the three questions as hypotheses, each with the measurement that tests it,
the outcome that would falsify it, and where the result is reported.

\paragraph{RQ1 (H1): the operator changes robustness to temporal shift} Under the null,
all architectures degrade equally and $\Delta\mathrm{AUC}$ differs only by optimisation
noise, so a random-split leaderboard would mis-rank designs only by a constant and
remain usable. We test with a Friedman omnibus over Run~A's nine variants on
$\Delta\mathrm{AUC}$ and on $\mathrm{AUC}_{2026}$, followed by corrected pairwise
comparisons over those variants and the flat control, and over a reproduction of
a published ensemble. A non-significant omnibus, or significant gaps not aligning with
any architectural property, would falsify H1 (Sections~\ref{sec:r-arch},
\ref{sec:r-mech} and \ref{sec:r-repro}).

\paragraph{RQ2 (H2): explanations degrade with accuracy} If a model's accuracy falls
across the boundary because it is attending to different, and wrong, parts of the
program, then the attribution profile should move with it. We compare per-concept
attribution profiles either side of the boundary with the Mann--Whitney $U$ test under
multiplicity correction. A significant shift supports H2; no detectable shift means the
accuracy loss is not accompanied by a change in what the model attends to. The
comparison is powered for a medium-to-large shift only, so we report a null result as
``no large shift detected'' (Section~\ref{sec:r-explain}).

\paragraph{RQ3 (H3): the RQ1 ranking carries design information} If H1 holds and the
separating property is real rather than an artifact of the architectures tested, an
architecture specified \emph{from} that property should perform like those it was
abstracted from. We build one in Run~B's $2\times2$ factorial after the temporal
comparison has run and evaluate it once under the same lock, with Run~C sweeping
regularizers over it; performance indistinguishable from the field's weaker half would
falsify H3. A pre-registered noise floor of 0.02 AUC decides in advance what counts as a
difference, which forces the comparison against the best searched operator to be
reported as a tie rather than a win (Section~\ref{sec:r-derived}).

\section{Experimental methodology}
\label{sec:method}
\label{sec:design}

Our object of study is an evaluation, not a detector. The study has five components,
listed in Table~\ref{tab:runs}: three training executions over the same temporal split,
which we call Run~A, Run~B and Run~C, plus a reproduction of the closest published
system and an attribution study over frozen checkpoints.

\begin{table}[!t]
  \centering
  \caption{The five experimental components of the study. Runs A, B and C are three
           separate training executions over the same temporal split.}
  \begin{tabular}{R{2.5cm}R{6.0cm}R{5.4cm}}
  \toprule
  \textbf{Component} & \textbf{What it does} & \textbf{What it answers} \\
  \midrule
  Run A & Nine graph neural network variants plus a flat-feature control, trained
          under the temporal protocol, five seeds each &
          Does the message-passing operator change robustness to temporal shift?
          (RQ1) \\
  Run B & A $2\times2$ backbone $\times$ readout factorial, the architecture derived
          from Run~A's finding, and three re-trained reference architectures &
          Is the Run~A finding prescriptive, and is the readout the mechanism
          behind it? (RQ3) \\
  Run C & A regularisation sweep over the derived architecture: five regulariser
          families, 15 configurations, 36 runs &
          Can the derived model's train--validation gap be closed, and does closing
          it help across the boundary? \\
  Reproduction & The attention-guided stacking ensemble of Shokouhinejad et al.\
          \cite{shokouhinejad2025stacked}, reimplemented to specification and run on
          this split &
          Can ensembling several operators substitute for choosing one robust
          operator? (RQ1) \\
  Attribution study & GNNExplainer \cite{ying2019gnnexplainer} applied to frozen
          checkpoints on both sides of the boundary &
          When a model degrades over time, does its explanation degrade with it?
          (RQ2) \\
  \bottomrule
  \end{tabular}
  \label{tab:runs}
\end{table}

The distinction between runs is substantive rather than a bookkeeping convenience. GPU
aggregation is not bit-deterministic, so two executions of the same architecture on the
same data, split and seeds differ slightly, and Run~C additionally consumes its random
number stream in a different order and so starts from different weights. We never pool
runs silently, and every comparison states which run it is made within.

\begin{figure}[!t]
  \centering
  \includegraphics[width=1.0\textwidth]{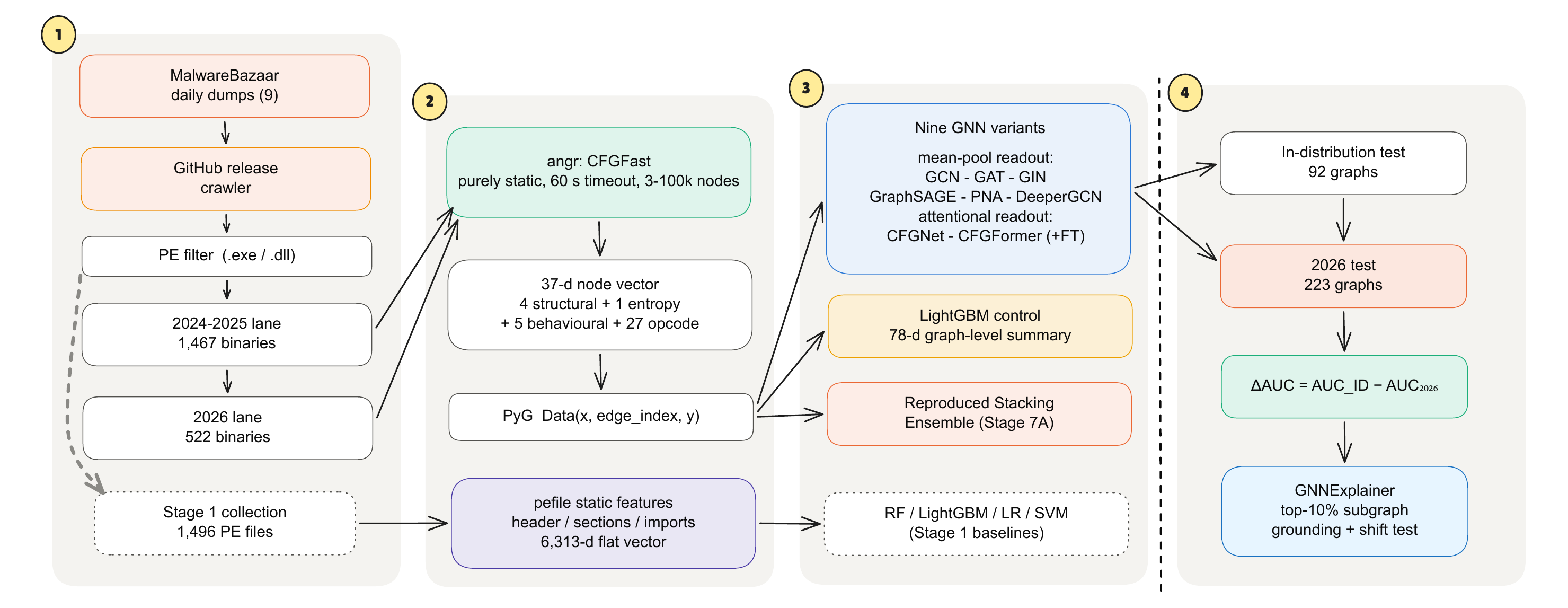}
  \caption{Overall workflow. Binaries are collected either side of a hard date
           boundary, converted to control flow graphs without execution, and scored
           under a protocol in which the later corpus is read exactly once.}
  \label{fig:workflow}
\end{figure}

Figure~\ref{fig:workflow} places these components in the end-to-end pipeline, with a
graph branch and a non-graph control branch scored under the same protocol.

\subsection{Corpora and the temporal boundary}
\label{sec:corpora}

\subsubsection{Why the corpora were collected rather than downloaded}
\label{sec:why-custom}

Using an established benchmark would make our results directly comparable to published
numbers, so departing from it requires justification. We impose four requirements at
once. R1, raw Portable Executable binaries, because control flow recovery disassembles
bytes and pre-extracted feature vectors cannot be turned back into a graph. R2, raw
binaries of \emph{both} classes, because the rank reversal (Section~\ref{sec:r-flat})
and the drift control (Section~\ref{sec:shift}) both need benign programs through the
identical extractor. R3, per-sample dates spanning a boundary recent enough that the
later side is genuinely unseen. And R4, obtainable at a workable scale, within the
compute and access available to an unaffiliated academic group.

Table~\ref{tab:datasets} records our review of the candidate corpora against these
requirements. No public corpus satisfies all four. The recurring pattern is that corpora
carrying dates do not release benign binaries, while corpora releasing both classes as
raw files carry no dates.

\begin{table}[!t]
  \centering
  \caption{Public malware corpora assessed against the four requirements above. No
           corpus meets all four: those carrying per-sample dates do not release
           benign binaries, and those releasing raw binaries of both classes carry
           no dates.}
  \setlength{\tabcolsep}{4pt}
  \begin{tabular}{R{2.4cm}R{3.0cm}R{2.4cm}cR{4.0cm}}
  \toprule
  \textbf{Corpus} & \textbf{Raw PE binaries?} & \textbf{Per-sample dates?} &
  \textbf{Fails} & \textbf{Reason} \\
  \midrule
  EMBER \cite{anderson2018ember} & No; feature vectors only & No & R1, R3 &
    No bytes, so no graph \\
  BIG 2015 \cite{ronen2018big2015} & No; header stripped, hex and disassembly
    listings & No & R1, R3 & As above \\
  BODMAS \cite{yang2021bodmas} & Malware only, request-gated & Partial,
    1999--2018 & R2, R3 & No benign binaries; window predates the study \\
  SOREL-20M \cite{harang2020sorel} & Malware only, $\approx$9.9M & Yes,
    first/last seen & R2, R4 & No benign binaries; $\approx$8\,TB \\
  DikeDataset \cite{dikedataset} & Yes, both classes & No & R3 &
    No dates, so no boundary \\
  PMML \cite{pmml2021} & Yes, both classes & No & R3 & As above \\
  VirusTotal, as used by MalGraph \cite{ling2022malgraph} & Yes, paid only & Yes &
    R4 & Bulk download needs a paid Enterprise API \\
  Drebin \cite{arp2014drebin} & Android APKs, request-gated & Partial,
    2010--2012 & R1, R2 & Not Windows PE; no benign release \\
  \midrule
  \textbf{This study} & Yes, both classes & Yes, per sample &
    --- & MalwareBazaar \cite{malwarebazaar} dumps; dated GitHub releases \\
  \bottomrule
  \end{tabular}
  \label{tab:datasets}
\end{table}

BODMAS \cite{yang2021bodmas} comes closest and its failure is instructive: it carries
per-sample timestamps but releases no benign binaries, so it supports a temporal split
over malware alone, not our experiment, since the benign class is simultaneously a model
input, the flat control's comparison set and the drift control. It is, for the same
reason, the constraint that shapes MalGraph's malware-only split.

The cost of that decision is stated here explicitly. A purpose-built corpus means our
absolute numbers are not comparable with figures published on EMBER, BODMAS or
SOREL-20M. Three things mitigate it: every comparison we make is \emph{internal}, with
all thirteen models on the same graphs, split and boundary; we reimplement the closest
published system on this corpus rather than comparing across corpora; and we release the
derived artifacts, so the corpus is auditable and reconstructible.

\subsubsection{The two dated corpora}

Two corpora carry the study (Table~\ref{tab:corpora}). A third, smaller collection is a
pilot on which we developed the pipeline and took two preliminary design decisions
(Section~\ref{sec:r-pilot}); it has no date structure, shares no binaries with the
others, and we never pool its results with theirs.

\begin{table}[!t]
  \centering
  \caption{The corpora. The two dated corpora were extracted in a single run and
           share an extractor version and configuration hash by construction.}
  \begin{tabular}{R{3.2cm}R{3.4cm}R{2.2cm}R{2.0cm}R{2.4cm}}
  \toprule
  \textbf{Corpus} & \textbf{Collection window} & \textbf{Binaries} &
  \textbf{Graphs} & \textbf{Class balance} \\
  \midrule
  Pilot & undated & 800 & 265 & 147 benign / 118 malware \\
  \texttt{cfg-2024-2025} (training) & 2024-01-01 to 2025-12-29 & 1,467 & 459 &
    166 benign / 293 malware \\
  \texttt{cfg-2026} (evaluation) & 2026-01-09 to 2026-07-22 & 522 & 223 &
    92 benign / 131 malware \\
  \bottomrule
  \end{tabular}
  \label{tab:corpora}
\end{table}

\begin{figure}[!t]
  \centering
  \includegraphics[width=1.0\textwidth]{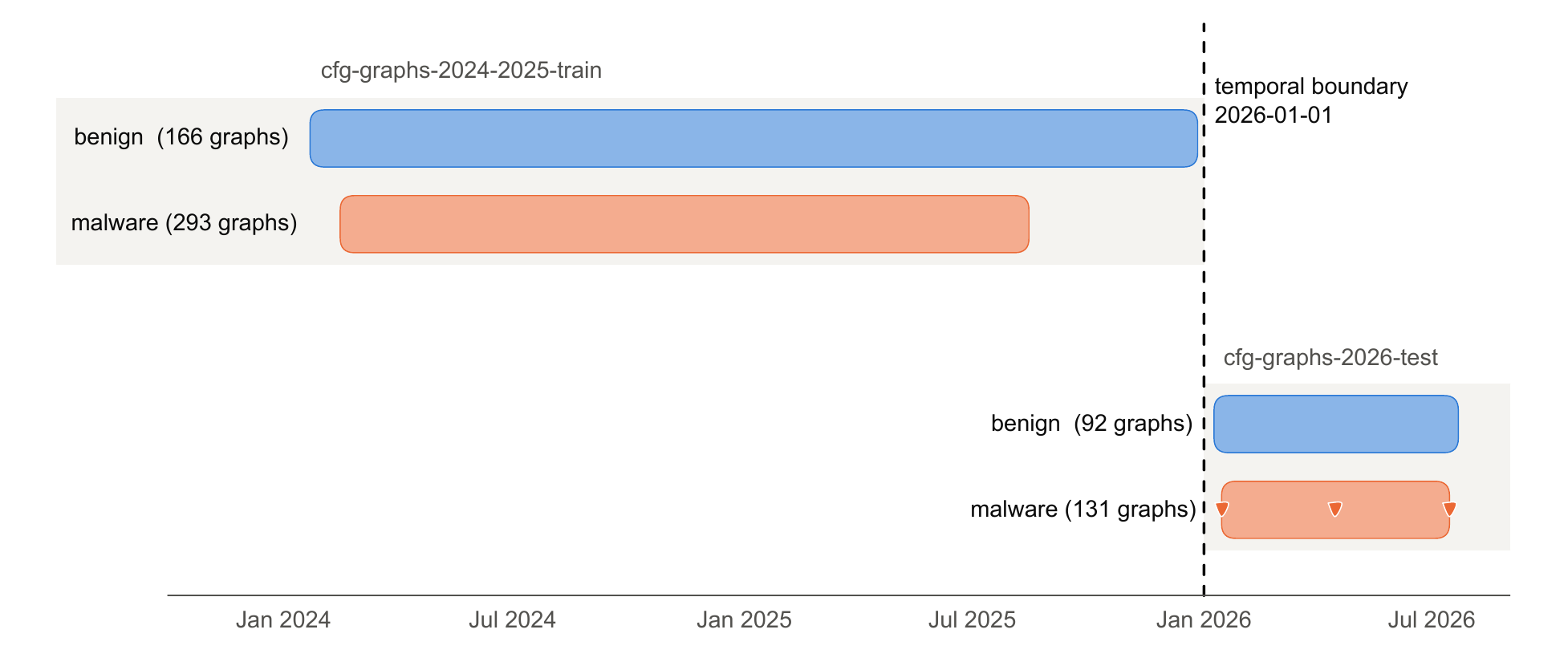}
  \caption{Timeline of the two dated corpora and the boundary that separates them.}
  \label{fig:timeline}
\end{figure}

We drew malware from daily MalwareBazaar \cite{malwarebazaar} dumps under the
corresponding date windows, restricted to Windows PE files, and crawled benign samples
from release artefacts of reputable GitHub projects over the same windows. Labels are
binary and source-determined, and every graph carries the SHA-256 digest and source date
of its binary (Figure~\ref{fig:timeline}).

We assert three leakage controls in code, because the whole study rests on the boundary
being real: duplicate SHA-256 digests are removed before any split is drawn; the
standardiser is fitted on training nodes only; and hyper-parameter search, early
stopping and threshold selection never read the 2026 corpus. A set intersection over
digests confirms zero overlap. Because we extracted both in a single pass they share
extractor version 1.0.0 and configuration hash \texttt{6bbe4a8649a76a3f}, with corpus
fingerprints \texttt{11d32afa64c19524} and \texttt{042c83c323203ba3}, so any difference
across the boundary is a property of the binaries, not of two differently configured
extractions.

\subsubsection{What the corpora cannot support}
\label{sec:data-limits}

Three limitations constrain how every later number reads. No family labels exist: the
field is empty for all 1,989 binaries, so every result is binary detection and the
covariate-shift versus concept-drift decomposition cannot be separated empirically. The
training corpus is not date-matched across classes (its malware ends 2025-08-15, its
benign samples run to 2025-12-27), though the 2026 boundary is clean on both sides. And
the evaluation corpus cannot support a within-test time series: four of its seven months
contain benign samples only, so any month-by-month breakdown is descriptive at best.

\subsection{Control flow graph extraction and node features}
\label{sec:cfg}

\begin{figure}[!t]
  \centering
  \includegraphics[width=1.0\textwidth]{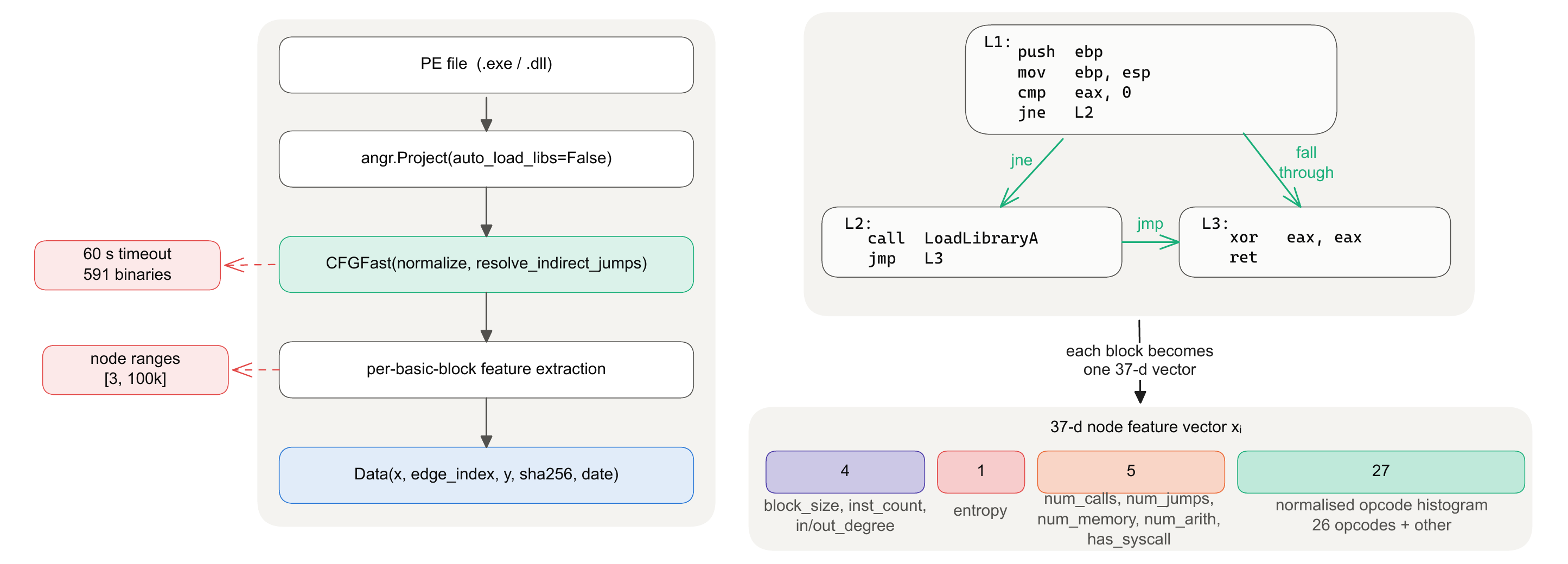}
  \caption{Static control flow graph extraction. No sample is executed; recovery is
           by disassembly and control-flow resolution only.}
  \label{fig:cfgpipe}
\end{figure}

Each binary becomes one directed graph of basic blocks and control-flow transitions.
Extraction is purely static. We recover graphs with angr's \texttt{CFGFast}
\cite{shoshitaishvili2016angr}, which disassembles and resolves control flow without
executing the binary, so we never run a sample, a safety property first and a
reproducibility property second, since a static pipeline has no sandbox-evasion
behaviour to account for. We load each binary with \texttt{auto\_load\_libs=False} so
library code is excluded, and run \texttt{CFGFast} with \texttt{normalize},
\texttt{resolve\_indirect\_jumps} and data-reference resolution enabled
(Figure~\ref{fig:cfgpipe}).

\subsubsection{Basic-block features}
\label{sec:nodefeat}

We describe every block by a 37-dimensional vector (Table~\ref{tab:nodefeat}). The
vector is deliberately inexpensive to compute, for a reason specific to this study: it
requires the disassembly of a single block and nothing else, so there is no vocabulary
to maintain and no out-of-vocabulary behaviour as the corpus moves forward in time. A
learned instruction embedding would add a second, uncontrolled source of temporal
degradation, the encoder ageing alongside the classifier, and would confound the
measurement this study is designed to make. Entropy is Shannon entropy over raw block
bytes and the opcode histogram is normalised by instruction count, so neither scales
with block size.

\begin{table}[!t]
  \centering
  \caption{The 37-dimensional basic-block feature vector.}
  \begin{tabular}{R{2.4cm}R{9.4cm}c}
  \toprule
  \textbf{Group} & \textbf{Features} & \textbf{Dim.} \\
  \midrule
  Structural & \texttt{block\_size}, \texttt{inst\_count}, \texttt{in\_degree},
    \texttt{out\_degree} & 4 \\
  Statistical & \texttt{entropy} (Shannon, over raw block bytes) & 1 \\
  Behavioural & \texttt{num\_calls}, \texttt{num\_jumps}, \texttt{num\_memory},
    \texttt{num\_arith}, \texttt{has\_syscall} & 5 \\
  Opcode & Normalised frequency over a fixed 26-opcode vocabulary (\texttt{mov},
    \texttt{push}, \texttt{pop}, \texttt{call}, \texttt{ret}, \texttt{jmp},
    \texttt{je}, \texttt{jne}, \texttt{jz}, \texttt{jnz}, \texttt{jl}, \texttt{jg},
    \texttt{jle}, \texttt{jge}, \texttt{add}, \texttt{sub}, \texttt{mul},
    \texttt{xor}, \texttt{and}, \texttt{or}, \texttt{lea}, \texttt{cmp},
    \texttt{test}, \texttt{nop}, \texttt{int}, \texttt{syscall}) plus an
    \texttt{other} bucket & 27 \\
  \midrule
  \textbf{Total} & & \textbf{37} \\
  \bottomrule
  \end{tabular}
  \label{tab:nodefeat}
\end{table}

Each surviving binary becomes a PyTorch Geometric \cite{fey2019pyg} \texttt{Data} object
carrying $\mathbf{X}\in\mathbb{R}^{N\times37}$, edge index $\mathbf{E}$, label $y$ and
provenance. Recovered graphs are large, at a median of 5,900 nodes on the dated corpora,
which is the scale at which the readout question of Section~\ref{sec:acfg} becomes
sharp.

\subsubsection{Extraction configuration, and why yield is a result}
\label{sec:extract-config}

The full configuration is \texttt{CFGFast} with \texttt{auto\_load\_libs} false and
\texttt{normalize}, \texttt{resolve\_indirect\_jumps} and \texttt{data\_references}
true; \texttt{cfg\_timeout\_sec} 60; a node-count window from 3 to 100,000; and the
37-dimensional node vector, at extractor version 1.0.0 and configuration hash
\texttt{6bbe4a8649a76a3f}. Two settings are methodological rather than incidental.

The per-binary timeout is binding: an earlier extractor raised a Python exception angr's
own handlers could swallow, producing extractions that ran far past their limit; ours
raises through the exception hierarchy backed by a parent-side
\texttt{SIGTERM}/\texttt{SIGKILL}. We set it to 60\,s rather than lower because angr's
analysis time tracks recovered graph size, not file size (about 0.3\,s at 121 nodes to
about 75\,s at 44,000), so the timeout is a \emph{sampling} decision and a low one
systematically discards the largest programs. Graphs below 3 nodes are stubs or packed
shells; graphs above 100,000 do not fit in GPU memory.

Because the timeout is a sampling decision, extraction yield is itself a result
(Section~\ref{sec:r-corpus}). The bias it introduces (over-representation of binaries
angr analyses quickly, so under-representation of heavily packed and obfuscated
programs) carries into every claim we make, and applies equally to other static
pipelines including MalGraph \cite{ling2022malgraph}.

\subsection{Model architectures}
\label{sec:archs}

\subsubsection{A shared backbone}

Comparing operators requires that nothing else vary, so every network shares one
skeleton and differences are attributable to the operator and readout rather than to
depth, width or head:
\begin{center}
\texttt{Input (37-d)} $\rightarrow$ $L \times$
\texttt{[Conv $\rightarrow$ BatchNorm $\rightarrow$ ReLU $\rightarrow$ Dropout]}
$\rightarrow$ \texttt{Readout} $\rightarrow$
\texttt{Linear(128$\rightarrow$64) $\rightarrow$ ReLU $\rightarrow$ Dropout
$\rightarrow$ Linear(64$\rightarrow$1)}
\end{center}
with $L=3$ layers and 128 hidden channels unless stated otherwise. Only \texttt{Conv}
and \texttt{Readout} vary between variants.

\subsubsection{Message-passing operators}

The five standard operators span the aggregation strategies in common use: smoothing,
edge-weighted attention, injective summation, separated self-and-neighbour updates, and
multi-aggregator combination. Writing $\mathbf{h}_i^{(l)}$ for node $i$ at layer $l$ and
$\mathcal{N}(i)$ for its neighbours,
\begin{align}
\text{GCN} \quad & \mathbf{h}_i^{(l+1)} = \sigma\!\Bigg(
  \sum_{j \in \mathcal{N}(i) \cup \{i\}} \frac{1}{\sqrt{d_i d_j}}\,
  \mathbf{W}^{(l)} \mathbf{h}_j^{(l)} \Bigg) \label{eq:gcn}\\
\text{GAT} \quad & \mathbf{h}_i^{(l+1)} = \Big\|_{k=1}^{K} \sigma\!\Bigg(
  \sum_{j \in \mathcal{N}(i)} \alpha_{ij}^{k}\, \mathbf{W}^{k}
  \mathbf{h}_j^{(l)} \Bigg) \label{eq:gat}\\
\text{GIN} \quad & \mathbf{h}_i^{(l+1)} = \mathrm{MLP}^{(l)}\!\Bigg(
  \big(1 + \epsilon^{(l)}\big)\mathbf{h}_i^{(l)} +
  \sum_{j \in \mathcal{N}(i)} \mathbf{h}_j^{(l)} \Bigg) \label{eq:gin}\\
\text{GraphSAGE} \quad & \mathbf{h}_i^{(l+1)} = \sigma\!\Bigg(
  \mathbf{W}_1 \mathbf{h}_i^{(l)} + \mathbf{W}_2 \cdot
  \operatorname*{mean}_{j \in \mathcal{N}(i)} \mathbf{h}_j^{(l)} \Bigg)
  \label{eq:sage}
\end{align}
Section~\ref{sec:rw-ops} introduced these operators; the settings that differ here are
GAT's $K=4$ heads of 32 channels and GIN's per-layer perceptron, which roughly doubles
GCN's parameter count.

PNA is included for a reason established by measurement rather than by preference. Our
preliminary study (Section~\ref{sec:r-pilot}) found GAT's learned attention to correlate
with out-degree at $\rho=0.84$, meaning a substantial part of the operator's capacity is
spent re-deriving connectivity the graph already encodes. PNA \cite{corso2020pna}
instead takes degree as an input, combining four aggregators with three degree scalers,
\begin{equation}
\bigoplus = \underbrace{\begin{bmatrix} I \\ S(D, \alpha{=}1) \\
  S(D, \alpha{=}{-}1) \end{bmatrix}}_{\text{scalers}} \otimes
  \underbrace{\begin{bmatrix} \mu \\ \sigma \\ \max \\ \min
  \end{bmatrix}}_{\text{aggregators}}
\label{eq:pna}
\end{equation}
and we run it with 4 towers and \texttt{divide\_input=True}. DeeperGCN
\cite{li2020deepergcn} uses GENConv with learnable softmax aggregation inside
\texttt{res+} residual blocks at 7 layers rather than 3; depth is intrinsic to that
architecture, so it is deliberately left unmatched to the others, and our comparison
against it is not parameter-matched.

\subsubsection{Two hypothesis-driven architectures}
\label{sec:custom-archs}

\begin{figure}[!t]
  \centering
  \includegraphics[width=1.0\textwidth]{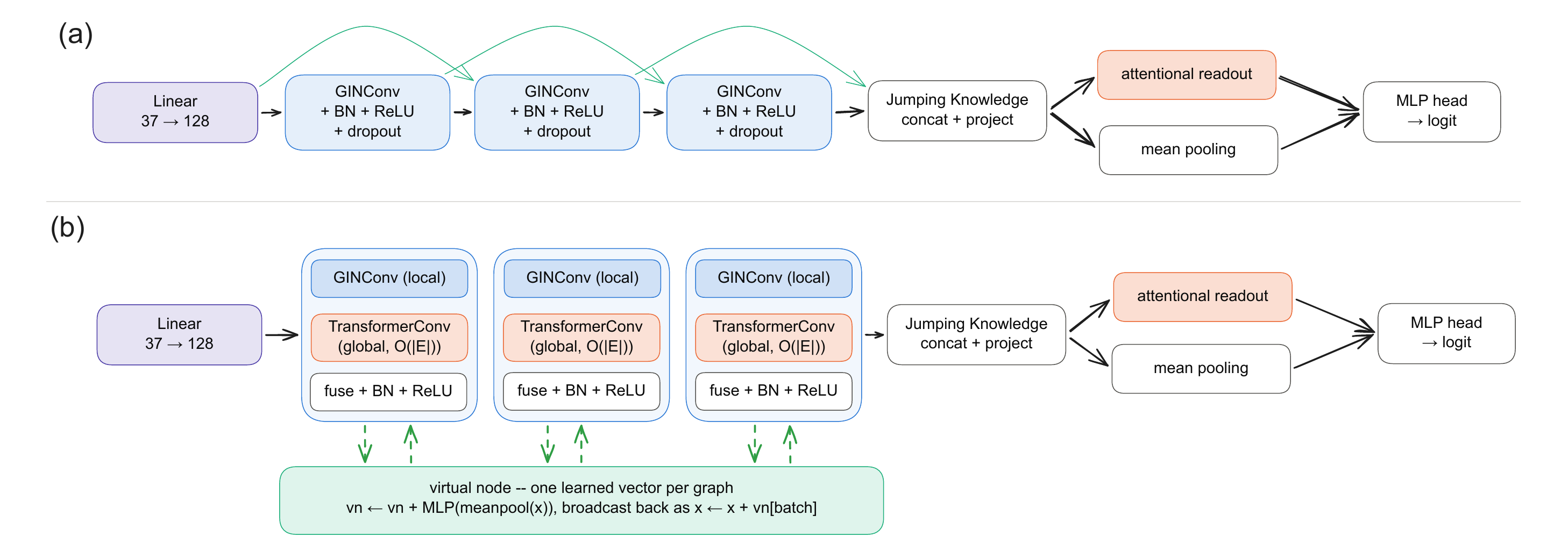}
  \caption{The two hypothesis-driven architectures. (a) CFGNet: a GIN
           backbone with residual connections, Jumping Knowledge and an attentional
           readout (178,821 parameters). (b) CFGFormer: a parallel local GIN and
           global \texttt{TransformerConv} update with a virtual node (574,853
           parameters).}
  \label{fig:customarch}
\end{figure}

We designed two architectures \emph{before} running any temporal comparison, each
testing a concrete hypothesis (Figure~\ref{fig:customarch}). Their design order matters:
they represent what a careful practitioner would build from domain intuition alone, and
Section~\ref{sec:r-derived} contrasts them with the architecture we designed afterwards
from measurement.

\paragraph{CFGNet} This architecture targets dilution: at a median of 5,900 nodes, if
maliciousness is concentrated in a small subgraph, mean pooling averages it away. It
keeps a GIN backbone but adds residual connections, Jumping Knowledge \cite{xu2018jk}
and an \emph{attentional readout} \cite{li2016ggnn} learning a per-node gate:
\begin{equation}
\mathbf{h}_G = \sum_{i} \mathrm{softmax}_i\big(g(\mathbf{h}_i)\big) \odot
  \mathbf{h}_i \quad \text{concatenated with} \quad \frac{1}{N}\sum_i \mathbf{h}_i
\label{eq:attnread}
\end{equation}

\paragraph{CFGFormer} This architecture tests whether long-range context helps: control
flow can carry a dependency between blocks many hops apart, which three rounds of local
message passing cannot reach. Each layer runs a local GIN update and a global
\texttt{TransformerConv} \cite{shi2021masked} update in parallel, in the spirit of
GraphGPS \cite{rampasek2022gps}, fused by a linear layer; a virtual node carries
whole-graph context at $O(|E|)$ cost rather than an $N\times N$ attention matrix, which
at these sizes would be prohibitive. It shares CFGNet's residuals, Jumping Knowledge and
attentional readout, so any difference is attributable to the encoder.

One CFGFormer variant tests whether drift robustness improves \emph{without} changing
the architecture, cheaper than redesign and worth ruling out first. Following GraphCL
\cite{you2020graphcl} we pretrain the backbone for 40 epochs under an NT-Xent loss with
$\tau=0.5$ on two augmented views per graph (feature masking and edge dropping, both at
rate 0.2); fine-tuning runs a 15-epoch frozen-backbone linear probe then a full unfreeze
at $0.1\times$ the discriminative learning rate. Pretraining reads only the structure of
the training and validation splits,
since pretraining on the evaluation corpus would leak the distribution the experiment
must generalise to.

\subsubsection{CFGNet-v2: an architecture derived from the result}
\label{sec:cfgnetv2}

We designed CFGNet and CFGFormer on intuition, and Run~A rejected both. We specified a
third only \emph{after} the temporal comparison had run, with every component chosen
because a measurement pointed at it. It is the instrument that tests H3: if the Run~A
finding is a property of operators rather than an accident of the architectures we
tested, an architecture assembled from it should behave like those it was abstracted
from.

CFGNet-v2 is a PNA backbone with residual connections, Jumping Knowledge and a plain
mean readout, at 3 layers and 297,857 parameters. Each choice is traceable: PNA because
Run~A found that operators building the graph vector by parametric aggregation survive
the boundary and PNA is the strongest, and because folding degree into the aggregation
answers the $\rho=0.84$ attention finding; residuals and Jumping Knowledge carried over
unchanged from CFGNet so the factorial can attribute effects to the two factors that
vary; and the mean readout as the direct reversal of CFGNet's one novelty. Width, depth,
dropout and head match the rest of the study.

\begin{figure}[!t]
  \centering
  \includegraphics[width=1.0\textwidth]{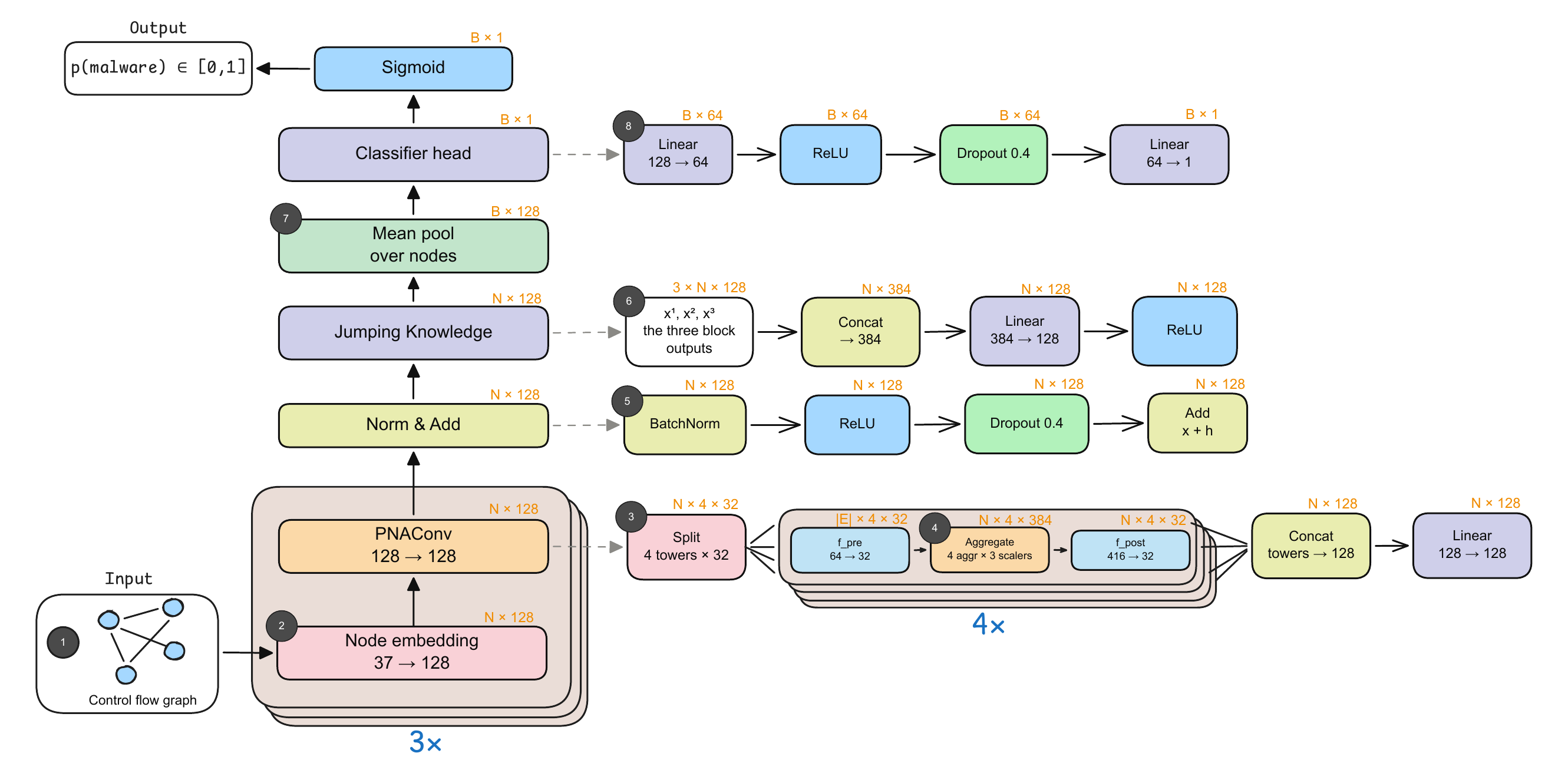}
  \caption{CFGNet-v2 in full, with the interior of one PNA layer expanded. The
           numbered badges key each block to its specification in
           Section~\ref{sec:cfgnetv2}.}
  \label{fig:cfgnetv2detail}
\end{figure}

Figure~\ref{fig:cfgnetv2detail} gives the architecture in full with one PNA layer
expanded. The three PNA layers hold 234,624 of the 297,857 parameters, Jumping Knowledge
49,280 and the head 8,321. Inside a layer, 128 channels split into four towers of 32;
each edge produces a message through a $64\rightarrow32$ map, reduced by all twelve
aggregator--scaler combinations, concatenated with the node's own state to give 416
channels and mapped $416\rightarrow32$; the towers concatenate back to 128 through a
final $128\rightarrow128$ layer, 78,208 parameters per layer.

\paragraph{Layer specification} The badges in Figure~\ref{fig:cfgnetv2detail} key each
block to its definition below, describing the built model: PyTorch Geometric's
\texttt{PNAConv} with four aggregators (\texttt{mean}, \texttt{max}, \texttt{min},
\texttt{std}), three degree scalers (\texttt{identity}, \texttt{amplification},
\texttt{attenuation}), four towers and a divided input.

{\footnotesize
\noindent\eqtag{1}~\textbf{Input.} A statically recovered graph $G=(V,E)$ with
$\mathbf{X}\in\mathbb{R}^{N\times 37}$ and $\mathbf{E}\in\mathbb{Z}^{2\times|E|}$,
$N=|V|$ (median $\approx 5{,}900$), each node vector concatenating the four
structural, one entropy, five behavioural and 27 opcode features of
Table~\ref{tab:nodefeat} and standardised by training-set statistics only,
$\tilde{\mathbf{x}}_i=(\mathbf{x}_i-\boldsymbol{\mu}_{\text{train}})
\oslash\boldsymbol{\sigma}_{\text{train}}$.

\noindent\eqtag{2}~\textbf{Node encoder.} One linear map with ReLU,
$\mathbf{x}^{(0)}_i=\operatorname{ReLU}(\mathbf{W}_{\text{enc}}
\tilde{\mathbf{x}}_i+\mathbf{b}_{\text{enc}})$ with
$\mathbf{W}_{\text{enc}}\in\mathbb{R}^{128\times 37}$, giving
$\mathbf{x}^{(0)}\in\mathbb{R}^{N\times 128}$.

\noindent\eqtag{3}~\textbf{PNAConv, four towers}, each aggregating independently over a
quarter of the channels:
$\mathbf{m}^{(t)}_{ij}=f^{(t)}_{\text{pre}}([\,\mathbf{x}^{(t)}_i\|
\mathbf{x}^{(t)}_j\,])$ and
$\mathbf{u}^{(t)}_i=f^{(t)}_{\text{post}}([\,\mathbf{x}^{(t)}_i\|
\bigoplus_{j\in\mathcal{N}(i)}\mathbf{m}^{(t)}_{ij}\,])$ for $t=1,\dots,4$, then
$\operatorname{PNAConv}(\mathbf{x})_i=\mathbf{W}_{\text{lin}}
[\,\mathbf{u}^{(1)}_i\|\cdots\|\mathbf{u}^{(4)}_i\,]$, with
$f_{\text{pre}}:\mathbb{R}^{64}\!\to\!\mathbb{R}^{32}$ and
$f_{\text{post}}:\mathbb{R}^{416}\!\to\!\mathbb{R}^{32}$.

\noindent\eqtag{4}~\textbf{Aggregation}, where degree enters as an input rather than
something the operator must re-derive:
$\bigoplus=[\,I, S(D,1), S(D,{-}1)\,]\otimes[\,\mu,\sigma,\max,\min\,]$, giving
$12\times32=384$ channels per tower, with
$S(D,\alpha)=(\log(d_i+1)/\delta)^{\alpha}$,
$\delta=|\mathcal{V}_{\text{train}}|^{-1}\sum_{i\in\mathcal{V}_{\text{train}}}
\log(d_i+1)$ and $d_i=|\mathcal{N}(i)|$.

\noindent\eqtag{5}~\textbf{Residual block}, $k=1,2,3$, so the node's own identity
survives three rounds of smoothing:
$\mathbf{h}^{(k)}=\operatorname{Drop}_{0.4}(\operatorname{ReLU}(\operatorname{BN}(
\operatorname{PNAConv}^{(k)}(\mathbf{x}^{(k-1)},\mathbf{E}))))$ and
$\mathbf{x}^{(k)}=\mathbf{x}^{(k-1)}+\mathbf{h}^{(k)}$.

\noindent\eqtag{6}~\textbf{Jumping Knowledge}, giving 1-, 2- and 3-hop views at once:
$\mathbf{z}_i=\operatorname{ReLU}(\mathbf{W}_{\text{jk}}[\,\mathbf{x}^{(1)}_i\|
\mathbf{x}^{(2)}_i\|\mathbf{x}^{(3)}_i\,])$ with
$\mathbf{W}_{\text{jk}}\in\mathbb{R}^{128\times384}$.

\noindent\eqtag{7}~\textbf{Readout}, the one component that defines v2, shown
against the attentional readout of Eq.~\eqref{eq:attnread} that it replaces:
\[
\begin{aligned}
\mathbf{h}_G&=\tfrac{1}{|V|}\textstyle\sum_{i\in V}\mathbf{z}_i\in\mathbb{R}^{128}
  &&\text{CFGNet-v2: plain mean}\\
\mathbf{h}^{\text{attn}}_G&=\Big[\,\textstyle\sum_{i\in V}a_i\mathbf{z}_i
  \,\Big\|\,\tfrac{1}{|V|}\textstyle\sum_{i\in V}\mathbf{z}_i\Big]
  \in\mathbb{R}^{256},\quad a=\operatorname{softmax}\big(g(\mathbf{z})\big)
  &&\text{CFGNet: removed in v2}
\end{aligned}
\]

\noindent\eqtag{8}~\textbf{Head and loss.} Two linear layers with dropout between
them, $\hat{y}=\mathbf{w}_2^{\top}\operatorname{Drop}_{0.4}(\operatorname{ReLU}
(\mathbf{W}_1\mathbf{h}_G+\mathbf{b}_1))+b_2$ with
$\mathbf{W}_1\in\mathbb{R}^{64\times 128}$ and $\mathbf{w}_2\in\mathbb{R}^{64}$,
trained under class-weighted binary cross-entropy at $w^{+}=0.567$.
}

\subsubsection{The factorial that breaks the confound}
\label{sec:factorial}

CFGNet differs from every drift-robust Run~A architecture in three ways at once
(backbone, residuals and Jumping Knowledge, and readout), so a readout claim resting on
the ranking alone rests on co-variation, which is the inference this study is designed
to avoid. Run~B adds two variants to complete a $2\times2$ over backbone and readout
with everything else fixed:

\begin{center}
\begin{tabular}{l|cc}
 & \textbf{attentional readout} & \textbf{mean readout} \\
\hline
\textbf{GIN backbone} & CFGNet & CFGNet (mean) \\
\textbf{PNA backbone} & CFGNet-v2 (attn) & CFGNet-v2 \\
\end{tabular}
\end{center}

The readout main effect is the mean column minus the attentional column, the backbone
effect is the PNA row minus the GIN row, and the interaction asks whether the readout
penalty depends on what it sits on. \texttt{CFGNet (mean)} is the control Run~A lacked:
CFGNet with exactly one component swapped. We declare two paired tests in advance as the
isolated readout test: \texttt{CFGNet (mean)} against CFGNet, and CFGNet-v2 against
CFGNet-v2 (attn). Declaring both arms in advance is what makes it possible to report, in
Section~\ref{sec:r-factorial}, that one fails.

Table~\ref{tab:archs} lists every variant in the study with its readout family and
parameter budget.

\begin{table}[!t]
  \centering
  \caption{Model variants, readout family and parameter budget. The nine above the
           rule form Run~A; the three below are the Run~B factorial, which also
           re-trains PNA, GraphSAGE and CFGNet as references.}
  \begin{tabular}{lR{5.6cm}cc}
  \toprule
  \textbf{Variant} & \textbf{Aggregation / readout} & \textbf{Depth} &
  \textbf{Params} \\
  \midrule
  GCN & degree-normalised sum / mean pool & 3 & 46,977 \\
  GAT & learned edge attention / mean pool & 3 & 47,745 \\
  GIN & sum / mean pool & 3 & 96,516 \\
  GraphSAGE & mean / mean pool & 3 & 84,481 \\
  PNA & 4 aggregators $\times$ 3 scalers / mean pool & 3 & 248,577 \\
  CFGNet & GIN + JK / attentional readout & 3 & 178,821 \\
  DeeperGCN & learnable softmax / mean pool & 7 & 477,320 \\
  CFGFormer & GIN + TransformerConv + virtual node /
    attentional readout & 3 & 574,853 \\
  CFGFormer+FT & as above, contrastively pretrained & 3 & 574,853 \\
  \midrule
  \textbf{CFGNet-v2} & PNA + JK / mean pool & 3 & 297,857 \\
  CFGNet-v2 (attn) & PNA + JK / attentional readout & 3 & 314,370 \\
  CFGNet (mean) & GIN + JK / mean pool & 3 & 162,308 \\
  \bottomrule
  \end{tabular}
  \label{tab:archs}
\end{table}

\subsubsection{A flat-feature control}
\label{sec:flat}

Message passing is only worth its cost if topology carries information the node features
alone do not, so we measure the graph models against a control seeing the same
information \emph{minus} topology. We collapse each graph to a 78-dimensional vector
(size and density statistics plus the mean and standard deviation of each of the 37 node
features) and classify it with LightGBM \cite{ke2017lightgbm} at 600 trees, learning
rate 0.05 and 31 leaves, over the same split and seeds. The control has every node
feature the graph models have and nothing about how blocks connect.
Section~\ref{sec:r-flat} reports that the answer depends entirely on which side of the
boundary the question is asked.

\subsection{Training configuration}
\label{sec:train}

All variants train under Table~\ref{tab:trainhp}, with a per-architecture search over
learning rate and dropout selected on validation AUC and five seeds (42--46) per
configuration; Run~B inherits it unchanged. The grid is four points rather than a full
cross product because the budget is a single GPU under a session limit, and more seeds
buys more against optimisation variance than more configurations.

\begin{table}[!t]
  \centering
  \caption{Shared training configuration for Runs A and B.}
  \begin{tabular}{R{4.6cm}R{7.6cm}}
  \toprule
  \textbf{Setting} & \textbf{Value} \\
  \midrule
  Hidden channels / layers & 128 / 3 (DeeperGCN: 7) \\
  Dropout & 0.4 (tuned per architecture) \\
  Loss & \texttt{BCEWithLogitsLoss}, $w^{+} = 0.567$ \\
  Optimiser & Adam, learning rate $10^{-3}$ (tuned), weight decay $10^{-4}$ \\
  Scheduler & \texttt{ReduceLROnPlateau}, factor 0.5, patience 8 \\
  Batch size & 16 \\
  Maximum epochs / early-stopping patience & 150 / 20 (on validation loss) \\
  Gradient clipping & 5.0 \\
  Learning rate and dropout grid & 4 points, not a full cross product:
    $(10^{-3}, 0.4)$, $(10^{-3}, 0.5)$, $(5{\times}10^{-4}, 0.4)$,
    $(5{\times}10^{-4}, 0.2)$ \\
  Seeds & 42, 43, 44, 45, 46 \\
  Node normalisation & \texttt{StandardScaler}, fitted on training nodes only \\
  \bottomrule
  \end{tabular}
  \label{tab:trainhp}
\end{table}

\subsection{Temporal evaluation protocol}
\label{sec:protocol}

We split the 459-graph training corpus 64/16/20, stratified at seed 42, into 293
training, 74 validation and 92 in-distribution test graphs, and hold the 223-graph 2026
corpus entirely separate. Class weighting uses $w^{+} = n_{\text{ben}}/n_{\text{mal}}
\approx 0.567$, malware being the majority class in this corpus.

We enforce the evaluation lock in code, which is what makes the 2026 figure an unbiased
estimate of deployment behaviour rather than a number optimised against: the
standardiser is fitted on training nodes only; early stopping and scheduling read
validation loss; the threshold is chosen on validation; and the 2026 corpus is loaded
once, after freezing. Run~B adds a further guard: after its search, an assertion
re-derives each selected configuration from the saved validation-AUC column alone and
fails the run if the two disagree.

Because the split is fixed across runs, the standard deviation we report isolates
optimisation variance, not split variance; Section~\ref{sec:stats} defines the paired
bootstrap covering the separate question of test-set sampling uncertainty.

\subsection{Statistical protocol}
\label{sec:stats}

We score binary classification with accuracy, precision, recall, F1, ROC-AUC,
average precision and the Matthews correlation coefficient, and report all seven for
the 2026 corpus in Table~\ref{tab:r-oper}. AUC is the ranking metric throughout, for
the reason given in Section~\ref{sec:temporal-gen}; we promote average precision to a
headline metric in Run~B, since the 2026 corpus is 59\% malware and precision-recall
area is the more informative summary at that balance.

Every comparison is \emph{paired over the same seeds}, since pairing removes the
variance that would otherwise dominate at $n=5$. The paired $t$-test is the primary
pairwise statistic; the Wilcoxon signed-rank test is reported alongside but never
treated as a failed check, since its floor at $n=5$ is $p=0.0625$. The Friedman test
\cite{friedman1937} is the omnibus for the architecture comparison, and families of
pairwise or per-concept tests are corrected with Benjamini--Hochberg
\cite{benjamini1995fdr}. The Nemenyi critical difference \cite{demsar2006} is reported
for completeness only: at $k=9$, $n=5$ it is 5.37, too wide to separate anything but top
from bottom. Stage-specific tests are Kolmogorov--Smirnov \cite{massey1951ks} for
distribution shift, Mann--Whitney $U$ for attribution shift, and Spearman $\rho$ with
Jaccard overlap for attention--degree correlation and cross-architecture agreement.
Where a sample is a handful of seeds we quote $p$ with Cohen's $d$; where roughly $10^5$
nodes are pooled we quote effect sizes without $p$-values, since at that $n$ any
$p$-value is uninformative.

\paragraph{Test-set sampling uncertainty} The seed standard deviation says nothing about
the 223 evaluation graphs being a finite sample. We quantify that with a paired
bootstrap over graphs: resampling the evaluation corpus with replacement $B=5{,}000$
times, we recompute each architecture's seed-averaged AUC on the resampled graphs, take
the difference between the two being compared, and report the 2.5th and 97.5th
percentiles. Pairing on the same resample is what makes the interval informative: the
marginal interval on any \emph{single} model's AUC is roughly $\pm0.05$ at this corpus
size, wide enough to swamp every gap in the study, while pairing removes the
graph-difficulty variation the two models share.

The 0.02 AUC noise floor declared in Section~\ref{sec:hypotheses} applies throughout: we
treat a smaller gap as noise and report it as a tie.

\subsection{Explainability protocol}
\label{sec:explain}

\subsubsection{Ruling out attention as an explanation channel}
\label{sec:attn}

If a GAT's attention weights highlighted malicious blocks, explainability would come
free with the operator. We test whether they do against a null model on the pilot
corpus. Since GAT normalizes attention over \emph{incoming} edges ($\sum_j
\alpha_{ij}=1$), the meaningful per-node quantity is the outgoing attention mass,
\begin{equation}
A_j = \sum_{i \in \mathcal{S}(j)} \alpha_{ij}
\label{eq:attnmass}
\end{equation}
extracted per edge across all layers, averaged over heads, self-loops excluded, and
reported both raw and per-edge-normalised so hub nodes cannot inflate the result.
The control that decides the outcome is a degree null model: we measure
concentration (normalised entropy, Gini, top-$k$ mass) on the attention
distribution \emph{and} on one derived from node degree alone. If attention is no more
concentrated than degree, the operator has learned nothing beyond topology. We fixed
three verdicts in advance: attention is \emph{localised} if normalised entropy $<0.85$
or top-5\% mass $>0.30$; it \emph{beats the null} if Gini excess $>0.02$ with paired
Wilcoxon $p<0.05$; and it is \emph{semantic} if any behavioural feature separates high-
from low-attention blocks at $|d|>0.3$ (Cohen's $d$ \cite{cohen1988} on raw features,
since with $\sim10^5$ pooled nodes any $p$-value is meaningless).

\subsubsection{Prediction-conditioned attribution}

Because the attention test returns a negative result (Section~\ref{sec:r-attn}), the
main study uses prediction-conditioned attribution. We apply GNNExplainer
\cite{ying2019gnnexplainer} to frozen Run~A checkpoints at seed 42 for PNA
(strongest on 2026) and GraphSAGE (most drift-robust), learning a soft node
and edge mask over 100 epochs at learning rate 0.01. These are the two architectures
Section~\ref{sec:r-arch} identifies as deployment candidates, so disagreement between
them is disagreement between viable alternatives. We skip graphs above 30,000 nodes,
explain fifteen graphs per (corpus $\times$ label) cell, 60 in total, and fix the
attributed subgraph in advance as the top 10\% of nodes by attribution score.

\paragraph{Behavioral grounding} The attributed nodes are compared against the rest of
the graph on ten behavioral concepts, two of them loop indicators, grouped as: API
activity, system interaction, obfuscation proxy, block complexity and tight loops.
Effect sizes are Cohen's $d$ on
raw features against a pre-registered $|d|>0.3$ threshold. \texttt{in\_degree} and
\texttt{out\_degree} are carried as explicit topology controls: grounding only on those
would mean the explainer has re-derived connectivity, which is exactly what the
attention test found the GAT is doing.

The shift test compares in-distribution and 2026 attribution profiles per
architecture and concept with Mann--Whitney $U$, rank-biserial effect size,
Benjamini--Hochberg corrected. At $n=30$ per side and 12 comparisons per architecture
only a medium-to-large shift would clear correction, so a null result means ``no large
shift detected''. Cross-architecture agreement compares top-10\% node sets by
Jaccard overlap and full attribution vectors by Spearman correlation, both against a
random top-10\% draw; without that baseline a small overlap could not be
distinguished from a meaningful one.

\subsection{Additional experiments}
\label{sec:addl}

\subsubsection{Learning curves, calibration and stratification}
\label{sec:supporting}

Three further experiments are carried out alongside Run~A. Learning curves ask whether the
drift ranking is an artifact of dataset size: we retrain each architecture on nested
subsets of the training split (25/50/75/100\%, that is 73/147/220/293 graphs) at three
seeds each against the fixed 2026 corpus, so that varying only the quantity of data
makes ``does capacity trade in-distribution accuracy for generalization'' a controlled
test rather than a correlation. A calibration analysis asks whether degradation
is a threshold problem rather than a ranking problem by re-thresholding every model
at its validation-optimal cut. Stratified analyses break degradation down by
graph-size tercile and by month, the latter descriptive only
(Section~\ref{sec:data-limits}).

\subsubsection{Testing whether the generalization gap can be regularized away}
\label{sec:gap}

CFGNet-v2 fits its 293 training graphs considerably harder than its 74 validation
graphs. The conventional reading is overfitting and the conventional response is to
regularise; Run~C tests both rather than asserting from a loss curve that the gap is
benign, asking two separable questions: whether a regulariser closes the gap, and
whether closing it helps across the boundary. We define the gap strictly, reloading the
frozen checkpoint and evaluating both terms in \texttt{eval()} mode:
\begin{equation}
\mathrm{gap} = \mathcal{L}_{\text{val}}^{\text{eval}} -
               \mathcal{L}_{\text{train}}^{\text{eval}}
\label{eq:gap}
\end{equation}
A training curve's running average is taken with dropout active and understates the gap
by roughly $1.3\times$; we record both readings.

We swept five regulariser families at two settings each (dropout (0.5, 0.6), weight
decay ($10^{-3}$, $5{\times}10^{-3}$), hidden width (96, 64), label smoothing (0.05,
0.10), and GraphCL-style augmentation applied during supervised training (0.1, 0.2)) on
the Run~B configuration and split, in three phases: an 11-run one-factor screen at one
seed, 15 runs at three seeds over the four best combinations plus a baseline, and 10
runs at five seeds over the winner plus a re-trained baseline. That is 15 configurations
across 36 runs, 8.8 hours on a single NVIDIA T4. Two rules were fixed first. The
\emph{selection rule} keeps configurations within 0.010 validation AUC of the best, then
takes the smallest mean $|\mathrm{gap}|$, absolute because a barely-fitted model can
post a negative gap. The \emph{success threshold} requires the gap to fall by at least
0.020 while 2026 AUC stays inside a $\pm0.02$ noise floor.

We declare one deviation because it changes how Run~C's control reads: Run~C builds the
classification head \emph{after} the convolutional stack, Run~B before. Architecture,
parameter count, data, split and seeds are identical, but the random number stream is
consumed in a different order so starting weights differ. Every Run~C configuration
shares one order, so comparisons within the sweep are clean; its baseline is a third
independent draw of CFGNet-v2 rather than a replication of Run~B's.

\subsubsection{Reproducing the attention-guided stacking ensemble}
\label{sec:repro}

RQ1 asks whether the operator matters; a natural objection is that it need not, if
operators can simply be ensembled. The attention-guided stacking ensemble of
Shokouhinejad et al.\ \cite{shokouhinejad2025stacked} makes exactly that move, so we
reimplemented it and ran it on our split. We chose it because it is the closest
published system in representation and task, and is built from three base learners
already in Run~A, so the comparison against single operators is internal rather than
cross-corpus.

We followed the published specification exactly: GCN, GIN and GAT base learners at 3
layers of 64 hidden units, dropout 0.2, cross-entropy loss, Adam with learning rate
$10^{-3}$ and weight decay $5{\times}10^{-4}$, 50 epochs, no batch normalisation and no
early stopping. Out-of-fold probabilities from 5-fold cross-validation are stacked into
$\mathbf{Y}=[\mathbf{Y}_1\|\mathbf{Y}_2\|\mathbf{Y}_3]$, and the attention meta-learner
computes
\begin{equation}
s_i = \mathbf{w}_i^{\top}\mathbf{y}_i + b_i, \qquad
\alpha_i = \mathrm{softmax}_i(s_i), \qquad
\Psi = \sum_{i=1}^{3} \alpha_i\, \mathbf{y}_i
\label{eq:meta}
\end{equation}
which feeds a $128\rightarrow64$ perceptron with dropout 0.2, trained for 100 epochs at
learning rate $10^{-3}$. We retrain base learners on the full 293-graph split for
inference and reproduce the paper's own ablation alongside it.

Two deviations are present in our implementations. First, the paper's meta-learner input is ambiguous between its formal
specification ($\Psi$) and its prose description (concatenation); we ran both readings,
labelled \texttt{psi} (the headline, matching the published equations) and
\texttt{concat}. Second, we replace their node representation (a rule-based
439-dimensional per-instruction encoding, aggregated per block then autoencoded to 64
dimensions) with our 37-dimensional block vector, because re-extracting theirs would
produce a different corpus under a different configuration hash and break the comparison
the reproduction exists to make. This is a reproduction of the paper's architecture and
protocol on our representation, never to be read as reproducing their headline accuracy.

\subsection{Implementation, reproducibility and safety}
\label{sec:impl}

We implement models in PyTorch \cite{paszke2019pytorch} and PyTorch Geometric
\cite{fey2019pyg}; control flow recovery uses angr \cite{shoshitaishvili2016angr}; PE
parsing uses \texttt{pefile} \cite{carrera2017pefile}; classical learners use
scikit-learn \cite{pedregosa2011sklearn} and LightGBM \cite{ke2017lightgbm}. All
training was performed on a single NVIDIA T4 under a hosted-notebook session limit,
which is why the study is organized into per-run wall-clock budgets with resumable
checkpointing. Extraction is CPU-bound and training GPU-bound, so the two run in
separate kernels, extraction versioned by extractor version and configuration hash.
Every run writes per-epoch history, the best checkpoint, SHA-256-keyed predictions and
a row in a runs table, and every figure rebuilds from those artifacts.

One reproducibility caveat is that angr's \texttt{CFGFast} is not bit-reproducible.
Re-extraction yields identical node counts, edge counts and features \emph{up to
permutation}, but a different node ordering. This is harmless for permutation-invariant
networks, but it means our fingerprint identifies the corpus rather than a byte-exact
set of graphs.

\section{Results}
\label{sec:results}

Two reading conventions apply throughout. Architectures are ranked on AUC rather than
F1 (Section~\ref{sec:temporal-gen}), and every $\pm$ is a standard deviation over five
seeds on one fixed split, so it measures optimisation variance; bracketed intervals are
the paired bootstrap over evaluation graphs (Section~\ref{sec:stats}).

\subsection{Representation and temporal shift}
\label{sec:r-shift}

\subsubsection{What the network propagates}
\label{sec:r-pilot}

A GCN on the 265-graph pilot corpus under a 169/43/53 split reaches AUC 0.895.
Replacing every node feature with a constant, leaving only topology, lowers AUC by
0.276 and binary F1 by 0.178, and the structure-only run's validation F1 reaches zero by
epoch 7. A 42-run ablation grid over three operators and four feature variants, with
channels zeroed rather than removed so that parameter counts stay constant, reproduces
this at every operator: structure-only F1 is 0.421, 0.208 and 0.571 for GCN, GAT and
GIN, against 0.709, 0.687 and 0.718 for the full vector. The opcode histogram alone
almost recovers the full model for GIN (AUC 0.882 against 0.866), indicating redundancy
within the 37 dimensions. The network therefore propagates basic-block content
(entropy, opcode mix, call density) along the control-flow structure rather than
classifying the structure itself.

On the pilot corpus only GIN separates from GAT on AUC (paired $t$, $p=0.017$,
$d=2.67$); GIN against GCN and GCN against GAT do not ($p=0.18$, $p=0.14$), and F1
separates none of them. At 265 graphs the supported ranking is GIN $\geq$ GCN $>$ GAT on
AUC with a three-way tie on F1, which does not survive to 459 graphs under a temporal
split.

\subsubsection{Extraction yield}
\label{sec:r-corpus}

Extraction produced 459 training graphs from 1,467 binaries (31\%) and 223 evaluation
graphs from 522 (43\%): 591 binaries exceeded the 60-second timeout and 367 recovered
graphs fell below the three-node floor. Per-class yield is 34\% and 30\%, so the loss
introduces no large class bias.

The recovered graphs are large: the pilot median is 2,678 nodes and 4,302 edges with a
tail to 97,119, and the dated corpora have a median near 5,900. Benign and malware
graphs are not separable from summary statistics. Mean node counts are 6,617 against
6,991, edges 9,554 against 10,511, mean degree 1.52 against 1.45 and entropy 2.56
against 2.57, none large enough to threshold on.

\subsubsection{The measured distribution shift is confounded}
\label{sec:r-drift}

Drift was measured separately in each class under the pre-registered rule of
Section~\ref{sec:shift}, with benign software as control (Table~\ref{tab:r-drift}).

\begin{table}[!t]
  \centering
  \caption{Structural drift between the 2024--2025 and 2026 corpora, measured
           separately in each class. The benign control drifts at least as much as
           malware.}
  \begin{tabular}{lccccc}
  \toprule
  \textbf{Class} & \textbf{Notable} & \textbf{Mean $D/D_{\text{crit}}$} &
  \textbf{Max $D$} & \textbf{Top feature} & \textbf{Domain AUC} \\
  \midrule
  Benign (control) & 32/40 & 1.27 & 0.401 & \texttt{mean\_op\_add} &
    \textbf{0.952 $\pm$ 0.026} \\
  Malware & 31/40 & 1.37 & 0.394 & \texttt{mean\_op\_xor} & 0.886 $\pm$ 0.015 \\
  \bottomrule
  \end{tabular}
  \label{tab:r-drift}
\end{table}

The corpora are distinct: 31 to 32 of 40 graph-level features are notable in each class,
and a domain classifier separates them at AUC 0.89 to 0.95. Malware drifts $1.08\times$
more than benign, and the domain classifier separates the corpora better on the benign
class (0.952) than on malware (0.886). The pre-registered verdict is therefore
\texttt{CONFOUNDED}: a toolchain, compiler or collection artefact explains the shift at
least as well as malware evolution. Everything from Section~\ref{sec:r-arch} onward
consequently concerns robustness to a measured distribution shift rather than malware
evolution.

\subsection{Architecture and temporal robustness}
\label{sec:r-arch}

\subsubsection{Operator and drift robustness}

\begin{figure}[!t]
  \centering
  \includegraphics[width=0.98\textwidth]{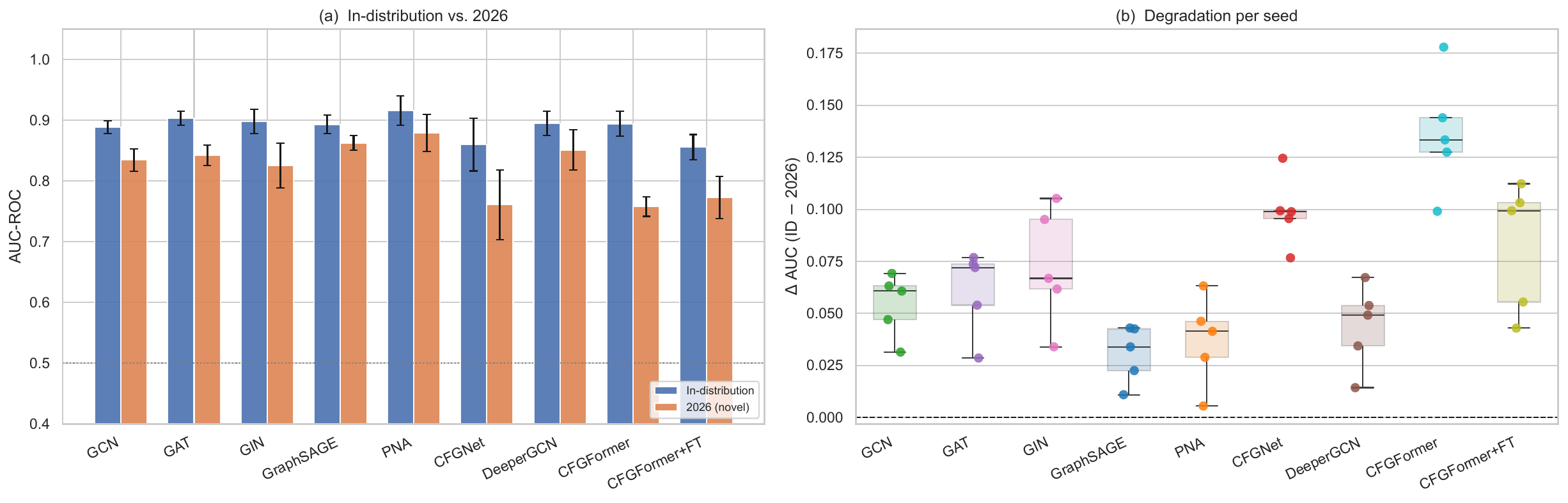}
  \caption{Temporal degradation across the nine variants of Run A. (a)
           In-distribution and 2026 AUC; (b) per-seed $\Delta$AUC. In-distribution
           performance does not predict performance on the later corpus.}
  \label{fig:r-degradation}
\end{figure}

Table~\ref{tab:r-runA} reports nine variants and the flat control, trained under the
evaluation lock and scored once on the 2026 corpus; Figure~\ref{fig:r-degradation} shows
the same field together with the per-seed spread in degradation. In-distribution
standing does not predict standing on the later corpus, and the spread in $\Delta$AUC is
wide, from 0.0306 for GraphSAGE to 0.1364 for CFGFormer.

\begin{table}[!t]
  \centering
  \caption{Run A: temporal degradation across nine variants and the flat control.
           Five seeds on one fixed split, so $\pm$ is optimisation variance. F1 here
           is at a fixed 0.5 cut; the validation-threshold operating point for the
           same models is in Table~\ref{tab:r-oper}. Measured training cost is
           reported in Section~\ref{sec:r-cost}.}
  \begin{tabular}{lccccc}
  \toprule
  \textbf{Architecture} & \textbf{ID AUC} & \textbf{2026 AUC} &
  \textbf{$\Delta$AUC} & \textbf{2026 F1@0.5} & \textbf{Params} \\
  \midrule
  PNA & 0.9162 $\pm$ 0.0243 & \textbf{0.8791 $\pm$ 0.0302} & $+$0.0371 &
    \textbf{0.824} & 248,577 \\
  GraphSAGE & 0.8933 $\pm$ 0.0154 & 0.8627 $\pm$ 0.0120 &
    \textbf{$+$0.0306} & 0.798 & 84,481 \\
  DeeperGCN & 0.8948 $\pm$ 0.0200 & 0.8510 $\pm$ 0.0330 & $+$0.0438 & 0.796 &
    477,320 \\
  GAT & 0.9033 $\pm$ 0.0116 & 0.8423 $\pm$ 0.0165 & $+$0.0610 & 0.653 & 47,745 \\
  GCN & 0.8889 $\pm$ 0.0106 & 0.8345 $\pm$ 0.0188 & $+$0.0543 & 0.685 & 46,977 \\
  GIN & 0.8980 $\pm$ 0.0202 & 0.8254 $\pm$ 0.0372 & $+$0.0726 & 0.724 & 96,516 \\
  CFGFormer+FT & 0.8560 $\pm$ 0.0205 & 0.7733 $\pm$ 0.0347 & $+$0.0827 & 0.744 &
    574,853 \\
  CFGNet & 0.8600 $\pm$ 0.0434 & 0.7610 $\pm$ 0.0575 & $+$0.0990 & 0.622 &
    178,821 \\
  CFGFormer & 0.8942 $\pm$ 0.0206 & 0.7578 $\pm$ 0.0160 & $+$0.1364 & 0.729 &
    574,853 \\
  \midrule
  \emph{LightGBM (flat)} & \emph{0.9320 $\pm$ 0.0027} & \emph{0.8074 $\pm$ 0.0049} &
    \emph{$+$0.1246} & \emph{0.785} & --- \\
  \bottomrule
  \end{tabular}
  \label{tab:r-runA}
\end{table}

A Friedman omnibus on $\Delta$AUC is significant in two independent executions with
separate hyper-parameter searches: an earlier six-architecture execution gives
$\chi^2=12.20$, $p=0.032$, and Run~A gives $\chi^2=27.41$, $p=0.0006$. The same tests on
2026 AUC give $p=0.012$ and $p=0.0003$. H1 is not rejected by any of the four.

Seven pairwise comparisons survive Benjamini--Hochberg correction in Run~A, and each
separates an aggregating architecture from one built around a learned attentional
readout: GraphSAGE, PNA, GAT, GCN and DeeperGCN each beat CFGFormer by 0.077 to 0.121
AUC ($p_{BH}$ between 0.0088 and 0.0343), PNA beats CFGNet by 0.118 ($p_{BH}=0.030$),
and GraphSAGE beats CFGFormer+FT by 0.089 ($p_{BH}=0.033$), with effect sizes from
$d=2.57$ to $d=7.41$. PNA and GraphSAGE are the top two on the 2026 corpus in both
executions and CFGNet is worst in both, despite independent tuning. A Nemenyi test is
uninformative at this scale, its critical difference being 5.37 at $k=9$ and $n=5$.

Under the paired bootstrap over the 223 evaluation graphs the largest gaps hold: PNA
over CFGNet is $+0.118$, $[+0.080, +0.158]$; GraphSAGE over CFGFormer $+0.105$,
$[+0.061, +0.153]$; and PNA over GIN $+0.054$, $[+0.016, +0.094]$. PNA over GraphSAGE
does not, at $+0.017$, $[-0.009, +0.044]$.

\paragraph{Operating point} Table~\ref{tab:r-oper} gives the full metric set on the 2026
corpus at each model's validation-selected threshold. PNA obtains its lead through
recall rather than precision, at 0.925 against 0.757, the most recall-skewed operating
point in the field. DeeperGCN, third on AUC, has the best accuracy, precision and
Matthews correlation of any model tested.

\begin{table}[!t]
  \centering
  \caption{Operating-point metrics on the 2026 corpus at each model's
           validation-selected threshold, averaged over five seeds. AUC and average
           precision are threshold-free and repeated from Table~\ref{tab:r-runA} for
           orientation. Ranking remains on AUC (Section~\ref{sec:temporal-gen}).}
  \begin{tabular}{lccccccc}
  \toprule
  \textbf{Architecture} & \textbf{AUC} & \textbf{Acc.} & \textbf{Prec.} &
  \textbf{Rec.} & \textbf{F1} & \textbf{MCC} & \textbf{AP} \\
  \midrule
  PNA & \textbf{0.879} & 0.781 & 0.757 & \textbf{0.925} & \textbf{0.832} &
    0.549 & 0.914 \\
  GraphSAGE & 0.863 & 0.744 & 0.755 & 0.855 & 0.799 & 0.466 & \textbf{0.915} \\
  DeeperGCN & 0.851 & \textbf{0.791} & \textbf{0.819} & 0.827 & 0.823 &
    \textbf{0.569} & 0.901 \\
  GAT & 0.842 & 0.770 & 0.781 & 0.853 & 0.813 & 0.522 & 0.881 \\
  GCN & 0.835 & 0.730 & 0.735 & 0.849 & 0.787 & 0.434 & 0.880 \\
  GIN & 0.825 & 0.747 & 0.758 & 0.853 & 0.799 & 0.473 & 0.886 \\
  CFGFormer+FT & 0.773 & 0.733 & 0.779 & 0.769 & 0.772 & 0.452 & 0.808 \\
  CFGNet & 0.761 & 0.724 & 0.738 & 0.821 & 0.777 & 0.423 & 0.777 \\
  CFGFormer & 0.758 & 0.705 & 0.767 & 0.721 & 0.740 & 0.403 & 0.789 \\
  \midrule
  \emph{LightGBM (flat)} & \emph{0.807} & \emph{0.740} & \emph{0.783} &
    \emph{0.771} & \emph{0.777} & \emph{0.465} & \emph{0.833} \\
  \bottomrule
  \end{tabular}
  \label{tab:r-oper}
\end{table}

\subsubsection{Graph structure pays off only out of distribution}
\label{sec:r-flat}

LightGBM on 78 graph-level summary statistics is the best in-distribution model in the
study at AUC 0.9320, above every graph network including PNA's 0.9162, and among the
worst on the 2026 corpus at 0.8074, with the second largest degradation measured
($\Delta=0.1246$). PNA beats it by $+0.072$ ($p=0.008$) and GraphSAGE by $+0.055$
($p=0.0015$). Under the paired bootstrap over evaluation graphs PNA's margin survives,
$[+0.023, +0.125]$, and GraphSAGE's does not, $[-0.002, +0.116]$. At least one graph
model therefore beats the topology-free control across the boundary by a margin
surviving both seed and graph resampling, having lost to it in distribution.

\subsubsection{Readout family and degradation}
\label{sec:r-readout}

Sorted by 2026 AUC, the three most drift-robust architectures of Table~\ref{tab:r-runA}
build the graph vector by parametric aggregation (GraphSAGE's mean, PNA's aggregator and
scaler mix, DeeperGCN's learnable softmax), and the two worst by learned attention:
CFGFormer ($\Delta=0.136$) and CFGNet ($\Delta=0.099$). CFGFormer is third best in
distribution at 0.894 and last on 2026 at 0.758. Depth, width and parameter count do not
follow the same ordering. CFGNet's design hypothesis, that an attentional readout
prevents mean pooling from diluting a small malicious subgraph, is not supported: it is
worst or second worst on 2026 in both executions. CFGNet and CFGFormer differ from the
drift-robust group in more than the readout, and Section~\ref{sec:r-factorial} isolates
that factor.

\subsection{Mechanism analysis}
\label{sec:r-mech}

\subsubsection{The degradation is ranking loss, not miscalibration}
\label{sec:r-calib}

\begin{table}[!t]
  \centering
  \caption{Score shift by class, and the F1 gap recovered by re-thresholding at the
           validation-optimal cut. Positive ``recovered'' means re-thresholding
           helped.}
  \begin{tabular}{lccccc}
  \toprule
  \textbf{Architecture} & \textbf{ID malware} & \textbf{2026 malware} &
  \textbf{$\Delta$F1@0.5} & \textbf{$\Delta$F1@thr} & \textbf{Recovered} \\
  \midrule
  PNA & 0.823 & 0.729 & 0.046 & 0.069 & $-$0.024 \\
  GraphSAGE & 0.755 & 0.645 & 0.061 & 0.088 & $-$0.027 \\
  GCN & 0.672 & 0.524 & 0.100 & 0.112 & $-$0.012 \\
  GIN & 0.754 & 0.621 & 0.099 & 0.096 & $+$0.004 \\
  GAT & 0.632 & 0.495 & 0.130 & 0.079 & $+$0.051 \\
  DeeperGCN & 0.781 & 0.660 & 0.086 & 0.071 & $+$0.015 \\
  CFGNet & 0.680 & 0.529 & 0.174 & 0.094 & $+$0.080 \\
  CFGFormer & 0.799 & 0.642 & 0.130 & 0.151 & $-$0.020 \\
  \bottomrule
  \end{tabular}
  \label{tab:r-shift}
\end{table}

Every architecture loses malware-side separation (Table~\ref{tab:r-shift}), PNA's mean
malware score falling from 0.823 to 0.729 while the benign side barely moves, and
re-thresholding at the validation-optimal cut recovers approximately none of the F1 gap,
going negative for PNA, GraphSAGE and CFGFormer. The models rank 2026 malware worse, so
the loss is not recoverable by post-hoc threshold adjustment, and the effect is
asymmetric between classes.

Two stratified observations accompany this. Small graphs, below 3,369 nodes, are hardest
on 2026 for every architecture (AUC 0.59 to 0.81, against 0.73 to 0.95 on the medium and
large terciles). A t-SNE \cite{vandermaaten2008tsne} projection of PNA's embeddings
shows 2026 graphs interleaved with the training distribution rather than forming a
separate cluster.

\subsubsection{Training set size}
\label{sec:r-lc}

\begin{figure}[!t]
  \centering
  \includegraphics[width=0.95\textwidth]{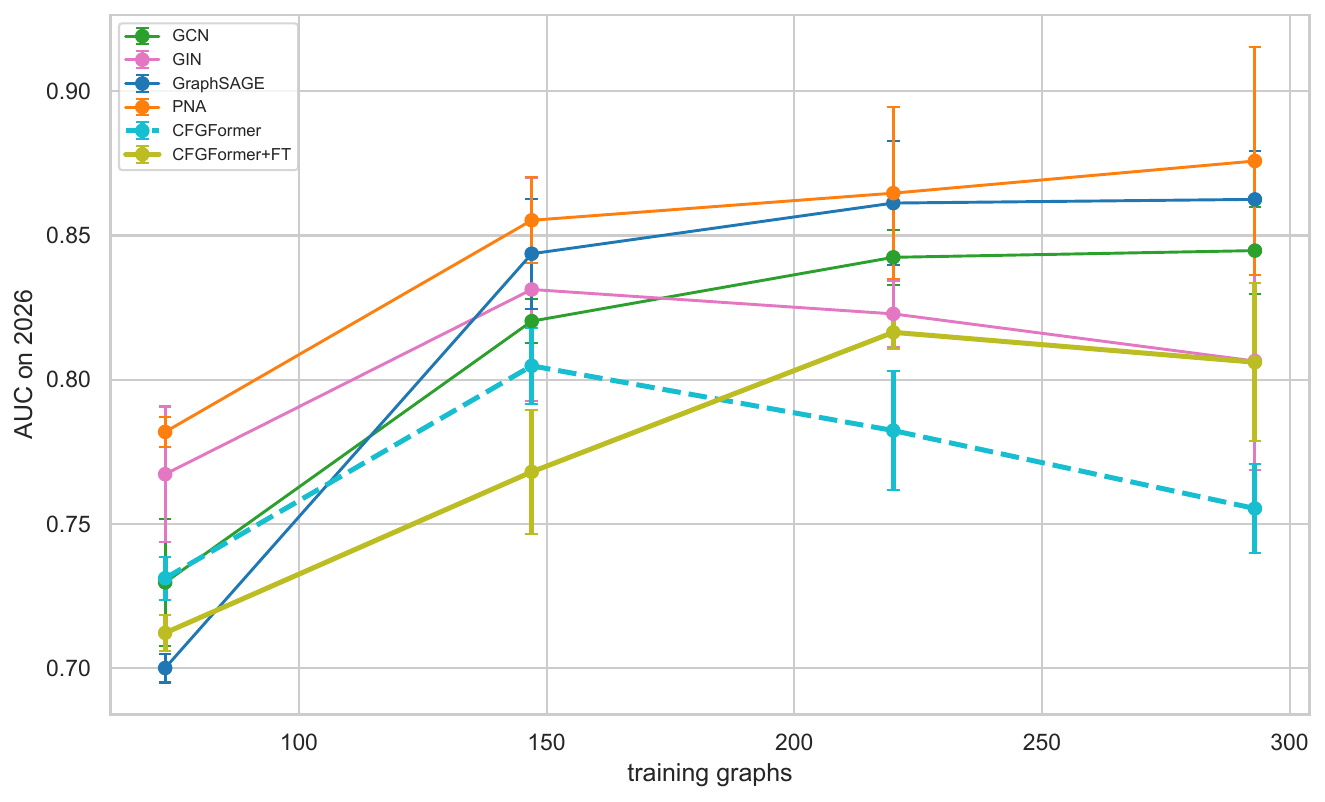}
  \caption{Learning curves against the fixed 2026 corpus. For the two
           highest-capacity models the in-distribution and out-of-distribution
           curves move in opposite directions.}
  \label{fig:r-lc}
\end{figure}

Going from 50\% to 100\% of the training data, CFGFormer gains 0.020 in-distribution AUC
and loses 0.050 on the 2026 corpus, and GIN gains 0.055 and loses 0.025; both peak on
2026 at half the available data (Figure~\ref{fig:r-lc}). GCN, PNA and GraphSAGE rise on
both axes, by $+0.024$,
$+0.021$ and $+0.019$ on 2026 respectively. For the two highest-capacity models, more
training data improves the in-distribution figure while lowering performance on the
later corpus.

Two observations bound how far a single-size comparison generalises: the spread across
architectures narrows only from 0.119 to 0.093 between the two ends, so the curves are
near-parallel rather than converging, and four of six architectures change rank between
the smallest and largest training set.

\subsubsection{Contrastive pretraining}
\label{sec:r-pretrain}

GraphCL-style pretraining \cite{you2020graphcl} was applied to CFGFormer, the
worst-degrading architecture, using only unlabelled training and validation structure.
Paired over five seeds, degradation falls from 0.1364 to 0.0827 ($d=-1.80$, $p=0.045$)
and in-distribution AUC from 0.8942 to 0.8560 ($d=-1.86$, $p=0.038$), while 2026 AUC
moves only from 0.7578 to 0.7733 ($p=0.51$) and average precision from 0.7886 to 0.8081
($p=0.43$). Pretraining also removes this architecture's learning-curve pathology.
CFGFormer+FT remains seventh of nine on 2026: the reduction in $\Delta$AUC is obtained
largely by lowering the in-distribution end, which is why $\Delta$AUC and 2026 AUC are
reported together throughout.

\subsubsection{Isolating the readout: the controlled factorial}
\label{sec:r-factorial}

Table~\ref{tab:r-runB} gives Run~B, the backbone $\times$ readout factorial with its
three references re-trained inside the run so that every paired test compares a single
execution.

\begin{table}[!t]
  \centering
  \caption{Run B: the backbone $\times$ readout factorial and its reference set,
           five seeds on the same split as Run A. All six rows come from one
           execution.}
  \begin{tabular}{lccccc}
  \toprule
  \textbf{Variant} & \textbf{ID AUC} & \textbf{2026 AUC} & \textbf{$\Delta$AUC} &
  \textbf{2026 AP} & \textbf{Params} \\
  \midrule
  \textbf{CFGNet-v2} & 0.9165 $\pm$ 0.0198 & \textbf{0.8769 $\pm$ 0.0202} &
    $+$0.0396 & \textbf{0.923} & 297,857 \\
  PNA & \textbf{0.9293 $\pm$ 0.0163} & 0.8734 $\pm$ 0.0190 & $+$0.0560 & 0.913 &
    248,577 \\
  GraphSAGE & 0.8933 $\pm$ 0.0151 & 0.8627 $\pm$ 0.0121 & \textbf{$+$0.0306} &
    0.915 & 84,481 \\
  CFGNet (mean) & 0.9040 $\pm$ 0.0115 & 0.8454 $\pm$ 0.0117 & $+$0.0586 & 0.895 &
    162,308 \\
  CFGNet-v2 (attn) & 0.8945 $\pm$ 0.0495 & 0.8401 $\pm$ 0.0100 & $+$0.0544 & 0.881 &
    314,370 \\
  CFGNet & 0.8722 $\pm$ 0.0581 & 0.7610 $\pm$ 0.0865 & $+$0.1112 & 0.793 & 178,821 \\
  \bottomrule
  \end{tabular}
  \label{tab:r-runB}
\end{table}

Re-training rather than reloading carries GPU non-determinism in scatter aggregation
worth roughly 0.006 AUC: GraphSAGE reproduces Run~A to four decimal places, while PNA's
2026 AUC moves from 0.8791 to 0.8734 and CFGNet's in-distribution AUC from 0.8600 to
0.8722. This is below the seed standard deviation of about 0.019 and changes no ranking,
and Table~\ref{tab:r-runB} is therefore read as self-contained.

\begin{table}[!t]
  \centering
  \caption{The $2\times2$: 2026 AUC and $\Delta$AUC by backbone and readout, from
           the four cell means of Table~\ref{tab:r-runB}.}
  \begin{tabular}{lcccc}
  \toprule
  & \multicolumn{2}{c}{\textbf{2026 AUC}} & \multicolumn{2}{c}{\textbf{$\Delta$AUC}} \\
  \textbf{Backbone} & \textbf{attentional} & \textbf{mean} & \textbf{attentional} &
  \textbf{mean} \\
  \midrule
  GIN & 0.7610 & 0.8454 & 0.1112 & 0.0586 \\
  PNA & 0.8401 & \textbf{0.8769} & 0.0544 & \textbf{0.0396} \\
  \midrule
  \multicolumn{5}{l}{\emph{Main effects on 2026 AUC:} readout $+0.0606$ $\cdot$
    backbone $+0.0553$ $\cdot$ interaction $-0.0476$} \\
  \multicolumn{5}{l}{\emph{Main effects on $\Delta$AUC:} readout $-0.0337$ $\cdot$
    backbone $-0.0379$ $\cdot$ interaction $+0.0378$} \\
  \bottomrule
  \end{tabular}
  \label{tab:r-factorial}
\end{table}

Table~\ref{tab:r-factorial} gives the four cell means and the derived effects. Every
cell points the same way: replacing the attentional readout with mean pooling improves
2026 AUC and reduces degradation on both backbones, and the readout main effect
($+0.061$) is the largest of the three.

The pre-registered paired tests support this on one arm only. On the GIN backbone,
\texttt{CFGNet (mean)} against CFGNet gains $+0.0844$ at $p=0.095$; on the PNA backbone,
CFGNet-v2 against CFGNet-v2 (attn) gains $+0.0368$ at $d=1.51$, $p=0.028$. Two features
bear on the GIN arm: CFGNet's 2026 AUC has a standard deviation of 0.087, four times
PNA's, and the interaction is negative ($-0.048$), the readout penalty being roughly
twice as large on GIN as on PNA. The readout is therefore one contributing factor of
two, comparable in size to the backbone and dependent on the backbone it sits on, and
the readout claim holds for whole architectures rather than for readouts alone.

\subsubsection{Cost}
\label{sec:r-cost}

Training cost was measured on a single NVIDIA T4. In seconds per epoch and peak GPU
memory in GiB: GraphSAGE 1.11 / 1.07, GIN 1.28 / 1.30, GCN 1.47 / 1.13, CFGNet 2.05 /
1.95, GAT 2.08 / 2.09, CFGFormer 5.67 / 3.75, DeeperGCN 9.41 / 9.22 and PNA 11.19 /
8.87. CFGNet-v2, measured in Run~B, runs at 11.6\,s per epoch; its peak memory was not
recorded. Parameter counts are in Table~\ref{tab:archs}.

PNA costs roughly ten times GraphSAGE per epoch and 8.3 times its peak GPU memory for
$+0.016$ AUC on the 2026 corpus, and degrades slightly more. Parameter count is not a
predictor of drift robustness: the two largest models in the field finish seventh and
ninth. Training cost is paid once, whereas the recurring cost is control flow recovery
at 0.3 to 75 seconds per binary (Section~\ref{sec:extract-config}), which cost 591 of
1,467 training binaries their place in the corpus.

\subsection{Evidence-driven architecture design}
\label{sec:r-derived}

CFGNet-v2 was specified from the Run~A finding rather than by trying candidates, and
evaluated once under the same lock (Table~\ref{tab:r-field}).

\begin{table}[!t]
  \centering
  \caption{CFGNet-v2 against the full field on 2026 AUC, paired over seeds 42--46.
           Benjamini--Hochberg correction is over the ten comparisons. Rows other
           than PNA, GraphSAGE and CFGNet are paired across runs and carry the
           additional cross-run variance discussed above.}
  \begin{tabular}{lccccc}
  \toprule
  \textbf{Comparator} & \textbf{Their 2026 AUC} & \textbf{Diff.} &
  \textbf{Cohen's $d$} & \textbf{$p$} & \textbf{$p_{BH}$} \\
  \midrule
  GCN & 0.8345 & $+$0.0424 & 7.52 & 0.0001 & \textbf{0.0007} \\
  CFGFormer & 0.7578 & $+$0.1191 & 6.22 & 0.0002 & \textbf{0.0008} \\
  CFGFormer+FT & 0.7733 & $+$0.1036 & 2.72 & 0.0037 & \textbf{0.0082} \\
  CFGNet & 0.7610 & $+$0.1159 & 2.65 & 0.0041 & \textbf{0.0082} \\
  \emph{LightGBM (flat)} & \emph{0.8074} & \emph{$+$0.0695} & \emph{2.84} &
    \emph{0.0032} & \textbf{\emph{0.0082}} \\
  GAT & 0.8423 & $+$0.0346 & 1.14 & 0.063 & 0.104 \\
  GIN & 0.8254 & $+$0.0515 & 0.93 & 0.105 & 0.150 \\
  GraphSAGE & 0.8627 & $+$0.0142 & 0.74 & 0.172 & 0.215 \\
  DeeperGCN & 0.8510 & $+$0.0259 & 0.55 & 0.284 & 0.315 \\
  PNA & 0.8791 & $-$0.0022 & $-$0.08 & 0.869 & 0.869 \\
  \bottomrule
  \end{tabular}
  \label{tab:r-field}
\end{table}

Five of ten comparisons survive correction: GCN, both CFGFormer variants, the original
CFGNet and the flat control. Against GAT, GIN, GraphSAGE and DeeperGCN the direction is
the same but the gap does not clear correction at five seeds. Against PNA the result is
a tie, at $-0.0022$ against Run~A's PNA ($p=0.87$) and $+0.0036$ against the PNA
re-trained inside Run~B ($p=0.79$), both inside the 0.02 noise floor. We claim no
improvement over PNA: an architecture specified from a measured finding matches the best
operator found by search, without search.

\subsubsection{Within-run advantages}

Three further readings separate CFGNet-v2 within Run~B. It has the highest floor, its
worst seed at 0.8538 against GraphSAGE's 0.8493 and PNA's 0.8443, and the highest best
seed at 0.9013. It has the highest average precision in its run, 0.9232 against 0.9151
and 0.9133, significant against GCN, GAT, both CFGFormers, CFGNet and its own
attentional arm. And it is the only model not dominated by GraphSAGE on the (2026 AUC,
$\Delta$AUC) plane. The independent draw of Section~\ref{sec:r-gap} removes two of the
three: the floor falls to 0.8504, level with GraphSAGE's 0.8501 across ten draws, and
average precision to 0.9072, below both. Non-dominance is also run-dependent, since on
Run~A numbers PNA leads on both axes. Only the floor separation from PNA survives.

CFGNet-v2 runs at 11.6\,s per epoch at 297,857 parameters, inheriting PNA's cost in
full, and its degradation ($\Delta=0.0396$) stays above GraphSAGE's 0.0306 ($p=0.53$).
Deriving the architecture from the finding produced the most accurate model on the later
corpus rather than the most drift-robust one. CFGNet-v2 (attn)'s search selected a
configuration with an outlier validation AUC of 0.9842, the likely source of its wide
spread ($\pm0.0495$); selection was validation-only and the leakage guard passed.

\subsubsection{The generalisation gap}
\label{sec:r-gap}

Measured strictly per Eq.~\eqref{eq:gap}, CFGNet-v2's train--validation gap at the
selected checkpoint is 0.122 with dropout disabled. Read off the loss curves it is 0.093
at the selected checkpoint, and 0.16 at the right-hand edge of the curves, where early
stopping discarded the model. All five seeds select checkpoints between epochs 41 and
86, against CFGNet's 25 to 78, one seed of which stopped at epoch 5 and whose 2026 AUC
spans 0.640 to 0.843.

\begin{table}[!t]
  \centering
  \caption{Run C: the selected regularisation against the Run B configuration,
           paired over seeds 42--46. The success threshold required the gap to fall
           by at least 0.020.}
  \begin{tabular}{lcccc}
  \toprule
  \textbf{Quantity} & \textbf{Baseline} & \textbf{Dropout 0.5} & \textbf{Diff.} &
  \textbf{$p$} \\
  \midrule
  Generalisation gap & 0.1216 & 0.1142 & $-$0.0074 & 0.45 \\
  Gap read off the loss curves & 0.0933 & 0.0774 & $-$0.0160 & 0.34 \\
  Validation AUC & 0.9572 & 0.9556 & $-$0.0016 & 0.73 \\
  ID AUC & 0.9236 & 0.9254 & $+$0.0018 & 0.64 \\
  2026 AUC & 0.8691 & 0.8709 & $+$0.0019 & 0.84 \\
  2026 average precision & 0.9072 & 0.9092 & $+$0.0020 & 0.86 \\
  $\Delta$AUC & 0.0545 & 0.0545 & $-$0.0000 & 0.99 \\
  \bottomrule
  \end{tabular}
  \label{tab:r-runC}
\end{table}

Run~C's pre-registered verdict is \texttt{GAP\_NOT\_REDUCED}. The selection rule chose
plain dropout 0.5 (Table~\ref{tab:r-runC}), which reduces the gap by 0.0074 against a
target of 0.020, at $p=0.45$. Configurations that tightened further paid in validation
AUC: dropout 0.6 reaches a gap of 0.065 for $-0.011$ validation AUC and was rejected by
the selection rule. Across all fifteen configurations, gap and 2026 AUC correlate at
$\rho=+0.28$, opposite in sign to the intuition and not significant ($p=0.32$).

The same pattern appears across architectures in Run~B, where the two tightest
train--validation gaps belong to the two weakest models on the 2026 corpus: CFGNet-v2
(attn) 0.014, CFGNet 0.019, PNA 0.062, CFGNet-v2 0.069, GraphSAGE 0.076 and CFGNet
(mean) 0.088.

Run~C's re-trained control is, by the declared initialisation-order deviation, a third
independent draw of CFGNet-v2, and it does not reproduce Run~B closely: 2026 AUC 0.8691
against 0.8769, in-distribution 0.9236 against 0.9165, $\Delta$AUC 0.0545 against
0.0396, and average precision 0.9072 against 0.9232. Neither AUC difference is
significant ($p=0.66$, $p=0.54$), but per-seed swings reach 0.05. Architecture, data,
split and seeds are identical and only the starting weights differ, and within-sweep
comparisons stay clean because every Run~C configuration shares one order. Across four
independent five-seed runs CFGNet-v2 and PNA remain bracketed within 0.010 AUC.

\subsection{Explanation stability and validity}
\label{sec:r-explain}

\subsubsection{Attention re-derives connectivity, not semantics}
\label{sec:r-attn}

Tested against the degree null model of Section~\ref{sec:attn} on the pilot corpus, the
pre-registered verdict for GAT attention is \texttt{DIFFUSE}. Mean normalised entropy
$H/\log N$ is 0.9589, and mean top-5\% attention mass is 0.1227, only $2.5\times$
uniform. Mean Gini excess over the degree null is $+0.1191$, which beats the null
($p = 2.4 \times 10^{-10}$), but attention mass correlates with out-degree at
$\rho = +0.8394$, and the Gini coefficient does not differ between malware and benign
graphs ($d = 0.285$, $p = 0.437$).

Profiling the top 5\% of blocks by attention mass, the two largest effects are
\texttt{num\_jumps} ($d=+0.734$) and \texttt{out\_degree} ($d=+0.714$), both
topological, while \texttt{op\_mov} ($-0.346$) and \texttt{op\_call} ($-0.341$) are
negative: high-attention blocks make fewer calls. Attention weights are consequently not
used as an explanation channel anywhere in this study, and PNA enters the model space as
the operator that takes degree as an input rather than re-deriving it. The negative
\texttt{op\_call} direction reappears in Section~\ref{sec:r-ground} from a different
method on a different corpus.

\subsubsection{Behavioural grounding is architecture-specific}
\label{sec:r-ground}

GNNExplainer was applied to frozen Run~A checkpoints for PNA, the strongest model on
2026, and GraphSAGE, the most drift-robust, over 60 graphs with the top-10\%
attribution threshold fixed in advance (Table~\ref{tab:r-ground}).

\begin{table}[!t]
  \centering
  \caption{Behavioural grounding of attributed subgraphs, as Cohen's $d$ between
           attributed nodes and the rest of the graph. Bold marks
           $|d| > 0.3$, the pre-registered threshold.}
  \begin{tabular}{lcc}
  \toprule
  \textbf{Concept} & \textbf{PNA} & \textbf{GraphSAGE} \\
  \midrule
  \texttt{num\_calls} & $-$0.047 & \textbf{$-$0.423} \\
  \texttt{op\_call} & $-$0.054 & $-$0.286 \\
  \texttt{entropy} & $+$0.007 & \textbf{$-$0.453} \\
  \texttt{inst\_count} & $-$0.007 & \textbf{$-$0.407} \\
  \texttt{block\_size} & $-$0.002 & \textbf{$-$0.411} \\
  \texttt{num\_jumps} & $+$0.117 & $-$0.039 \\
  \texttt{has\_syscall}, \texttt{op\_int} & 0.000 & 0.000 \\
  \texttt{self\_loop}, \texttt{reciprocal} & $-$0.003 / $+$0.013 &
    $-$0.018 / $-$0.036 \\
  \midrule
  \texttt{in\_degree} \emph{(topology control)} & $+$0.215 & $-$0.027 \\
  \texttt{out\_degree} \emph{(topology control)} & $+$0.152 & $-$0.304 \\
  \midrule
  $\rho$(attribution, out-degree) & 0.182 & $-$0.115 \\
  Gini excess over degree null & $-$0.075 & $-$0.108 \\
  \bottomrule
  \end{tabular}
  \label{tab:r-ground}
\end{table}

The two architectures differ completely. PNA's attributions ground on no behavioural
concept: every concept sits below $|d|=0.22$, and correlation with out-degree is 0.18,
far below the 0.84 measured for GAT attention, so the explainer is not re-deriving
connectivity either.

GraphSAGE grounds strongly and in the opposite direction to the intuitive hypothesis.
Its attributed nodes have lower call density, entropy, instruction count and block size
than the rest of the graph, with $|d|$ between 0.3 and 0.45, corresponding to small,
quiet blocks such as dispatchers, thunks and jump-table stubs rather than the busy
API-heavy blocks an analyst would nominate. The direction agrees with the negative
\texttt{op\_call} result obtained independently from GAT attention on a different
corpus. GraphSAGE's grounding is near zero on small graphs and strong on medium and
large ones ($d \approx -0.4$ to $-0.59$), while PNA's stays weak across all terciles.

\subsubsection{Attributions do not shift across the boundary}

\begin{figure}[!t]
  \centering
  \includegraphics[width=0.98\textwidth]{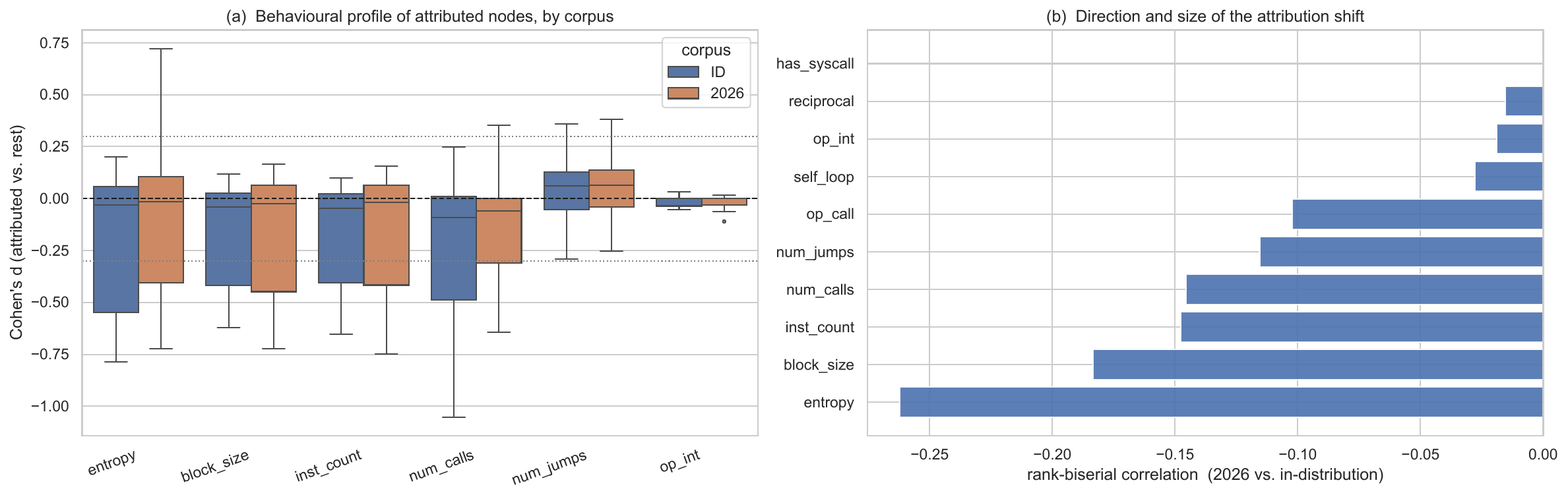}
  \caption{Attribution profiles either side of the temporal boundary. No behavioural
           concept and no topology control shifts significantly after correction.}
  \label{fig:r-shift}
\end{figure}

Zero of 20 behavioural comparisons and zero of 4 topology controls survive
Benjamini--Hochberg correction under Mann--Whitney $U$ (Figure~\ref{fig:r-shift}). The
largest raw effect is PNA on entropy (rank-biserial $-0.347$, $p=0.022$), which does not
survive correction ($p_{BH}=0.258$). The accuracy drop of Section~\ref{sec:r-arch} is
therefore not accompanied by a detectable change in what the models attend to. At
$n=30$ per side and 12 comparisons per architecture the test is powered for a
medium-to-large shift only, so the result is an absence of detected shift rather than
demonstrated invariance.

\subsubsection{The two architectures agree at chance}

Jaccard overlap between the two models' top-10\% node sets is 0.0558 in distribution and
0.0549 on 2026, against a random-draw expectation of 0.0526, and Spearman correlation
between the full attribution vectors is effectively zero, at $-0.030$ and $-0.033$. Two
architectures within 0.016 AUC of each other on the same graphs select essentially
disjoint evidence for the same predictions.

\subsection{Reproduced stacking ensemble}
\label{sec:r-support}
\label{sec:r-repro}

Table~\ref{tab:r-repro} reports our reimplementation of the attention-guided stacking
ensemble of Shokouhinejad et al.\ \cite{shokouhinejad2025stacked} on the temporal split.

\begin{table}[!t]
  \centering
  \caption{Reproduction of the attention-guided stacking ensemble on the temporal
           split, with two Run A architectures for reference.}
  \begin{tabular}{lcccc}
  \toprule
  \textbf{Model} & \textbf{ID AUC} & \textbf{2026 AUC} & \textbf{$\Delta$AUC} &
  \textbf{2026 F1} \\
  \midrule
  GCN & 0.855 $\pm$ 0.004 & 0.779 $\pm$ 0.015 & 0.076 & 0.718 \\
  GIN & 0.857 $\pm$ 0.008 & 0.789 $\pm$ 0.015 & 0.068 & 0.777 \\
  GAT & 0.863 $\pm$ 0.002 & \textbf{0.803 $\pm$ 0.011} & 0.061 & 0.720 \\
  Average Ensemble & 0.863 $\pm$ 0.004 & 0.804 $\pm$ 0.008 & 0.059 & 0.752 \\
  Stacking Ensemble (psi) & 0.863 $\pm$ 0.004 & 0.804 $\pm$ 0.011 & 0.059 &
    0.747 \\
  Stacking Ensemble (concat) & 0.865 $\pm$ 0.002 & 0.796 $\pm$ 0.008 & 0.069 &
    0.786 \\
  \midrule
  \emph{PNA (Run A)} & \emph{0.916} & \emph{\textbf{0.879}} & \emph{0.037} &
    \emph{0.824} \\
  \emph{GraphSAGE (Run A)} & \emph{0.893} & \emph{0.863} & \emph{0.031} &
    \emph{0.798} \\
  \bottomrule
  \end{tabular}
  \label{tab:r-repro}
\end{table}

The reproduction is faithful in distribution. Base-learner accuracies of 0.789, 0.807
and 0.794 for GCN, GIN and GAT land two to four points below the 0.8313, 0.8253 and
0.8373 those authors report, the expected direction and magnitude for a 37-dimensional
block vector in place of their autoencoder-reduced encoding. These figures do not
reproduce that paper's headline 0.8614 accuracy at 0.939 AUC on its own corpus.

The paper's central claim does not reproduce. The stacking ensemble beats its best base
learner by $+0.0015$ AUC ($p=0.62$) and the Average Ensemble by $+0.0005$ ($p=0.81$). It
does beat the alternative \texttt{concat} reading of its own specification ($+0.0085$,
$d=0.92$, $p=0.023$). Against single operators the ensemble loses to Run~A's PNA by
$-0.075$ AUC ($d=-3.32$, $p=0.004$) and to a single GraphSAGE by $-0.059$ ($d=-5.19$,
$p=0.0004$).

The learned meta-weights are $\alpha=0.438/0.297/0.266$ for GIN/GAT/GCN in distribution
and $0.436/0.298/0.266$ on 2026, almost input-invariant across a boundary that moves
base-learner accuracy by six to eight points. GIN carries the highest weight despite
being second worst on 2026, and GAT the lowest despite being best.

\section{Discussion}
\label{sec:discussion}

\subsection{RQ1: does the operator change temporal robustness?}
\label{sec:d-rq1}

The operator does change temporal robustness, and the change is measurable. The Friedman
omnibus on temporal degradation is significant in two independent executions with
separate hyper-parameter searches, seven pairwise gaps on out-of-distribution AUC
survive Benjamini--Hochberg correction, the largest also survive resampling the
evaluation graphs, and the identity of the top two architectures and the bottom one is
stable across executions despite independent tuning. H1 survives every test we set it.
Five further measurements stand behind that answer, in descending order of weight.

Architectures that aggregate transfer; those built around learned attention do not. This
continues Section~\ref{sec:r-attn}: a GAT's attention re-derives out-degree rather than
discovering semantics, and given capacity to learn beyond degree, what it learns is
corpus-specific. The ranking, CFGFormer's reversal and CFGNet's failure all point the
same way. Our wording is deliberate: the factorial test shows the readout effect largest
of the three but significant only on the PNA backbone, with a substantial negative
interaction, so the claim holds for \emph{whole architectures}, not readouts in
isolation.

Graph structure only pays for itself out of distribution. A flat gradient-boosted model
given every node feature but no topology is the best in-distribution model in the study
and among the worst across the boundary, and PNA's margin over it survives both seed and
graph resampling. This is the paper's principal methodological result, and it is
independent of which GNN wins: the choice of evaluation protocol, not merely of metric,
determines which \emph{family} of models a study recommends.

Capacity trades in-distribution accuracy for temporal generalisation. The learning
curves are closer to a controlled test than a correlation across architectures, since
only the quantity of data varies, and they answer what practitioners ask first: more
data does \emph{not} always help, and for the two highest-capacity models it actively
hurts.

The degradation is ranking loss, not miscalibration. Re-thresholding recovers
approximately none of the F1 gap and goes negative for three architectures, so the
operator choice cannot be deferred to post-hoc recalibration.

Ensembling does not substitute for choosing one robust operator. The reproduced stacking
ensemble loses on the later corpus to both a single PNA and a single GraphSAGE, and its
meta-learner is almost input-invariant across the boundary: independent support for the
same conclusion, the stronger for coming from an ensemble designed by other authors for
a different purpose.

\subsection{RQ2: do explanations survive the same shift?}
\label{sec:d-rq2}

Explanations do survive the shift, within the power the sample allows, but with an
unanticipated complication. No behavioural concept and no topology control shifts
significantly across the boundary, so the accuracy drop is not accompanied by a
detectable change in what the models attend to: they keep looking at similar structure
and get it wrong more often. One corollary follows immediately: an attribution cannot
serve as a drift alarm, since it looks the same on both sides of a boundary costing
several points of AUC.

The complication is that \emph{explanation validity is a property of the architecture
rather than of the data or the task}. PNA grounds in no behavioural concept at all;
GraphSAGE grounds strongly but opposite to the intuitive hypothesis, selecting small,
quiet blocks rather than busy API-heavy ones; and the two agree at chance level. Our
conclusion is:

\begin{quote}
GNN explanation validity in this domain is architecture-specific, and the
operator that scores highest on the later corpus is the one hardest to explain
locally.
\end{quote}

Three measurements carry this, none the one we expected the question to turn on:
Section~\ref{sec:r-attn} first had to \emph{rule out attention as an explanation
channel}, and Section~\ref{sec:r-ground} then finds \emph{no shift} but \emph{no shared
grounding}. The practical consequence is a caution rather than a technique: an
attribution shown to an analyst, an auditor or a court is a property of the architecture
at least as much as of the binary.

\subsection{RQ3: is the drift finding prescriptive?}
\label{sec:d-rq3}

If the finding of Section~\ref{sec:d-rq1} is real rather than an artefact of the
architectures tested, an architecture specified \emph{from} it should behave like those
it was abstracted from, a falsifiable prediction that Section~\ref{sec:r-derived} tests.

CFGNet-v2 matches the best operator at a tie rather than a win ($\pm0.004$ AUC across
two independent PNA draws, $p \geq 0.79$), and beats the flat control, GCN, both
CFGFormers and the original CFGNet under correction. We claim no improvement over PNA.
The value of the result is not the margin but the provenance: we derived the
architecture rather than discovering it, and the two architectures we designed on
intuition finished last and second to last while the one we designed on evidence
finished first. A ranking that carries enough information to specify a competitive new
design is a different kind of object from a ranking that merely orders existing ones.

Two further results bound the answer, both cutting against the architecture. The three
readings on which CFGNet-v2 leads (highest floor, highest average precision,
non-dominance) hold only \emph{within the run in which we measured them}; an independent
draw leaves only the separation from PNA standing. And the regularisation sweep returns
\texttt{GAP\_NOT\_REDUCED}: the 0.122 train--validation gap resists five regulariser
families, and closing it would not have helped, since gap and out-of-distribution AUC
are uncorrelated across all fifteen configurations and the two tightest gaps in the
factorial belong to the two weakest models. A large train--validation gap is not the
pathology it appears to be on this corpus.

What the answer does not extend to: CFGNet-v2's degradation stays above GraphSAGE's, it
inherits PNA's cost in full, and it is the most \emph{accurate} model on the later
corpus rather than the most \emph{robust}. Deriving from the drift finding produced
accuracy, not robustness.

\subsection{Practical implications}
\label{sec:d-implications}

\subsubsection{Which architecture to deploy}
\label{sec:d-tension}

The three answers do not point at the same architecture, and we report the disagreement
rather than reconciling it: Section~\ref{sec:d-rq1} recommends GraphSAGE on robustness
per unit of compute; Section~\ref{sec:d-rq3} makes CFGNet-v2 the most accurate model on
the later corpus; and Section~\ref{sec:d-rq2} notes that CFGNet-v2 and PNA share a
backbone whose decisions no sparse subgraph explains, while GraphSAGE's can be,
counter-intuitively.

We recommend GraphSAGE. It sits within 0.014 AUC of CFGNet-v2 on the later corpus, has
the lowest degradation of anything we tested, runs on an eighth of PNA's peak GPU
memory, and produces attributions that are at least characterisable. Its tenfold
training-time advantage matters less than it appears: static control flow recovery costs
0.3 to 75 seconds per binary and dominates the recurring cost by orders of magnitude, so
seconds per epoch bear on retraining cadence, not detection throughput. The case rests
on degradation and memory. The counter-intuitive direction of its attributions also
yields a testable hypothesis: if the signal lies in control-flow plumbing (dispatchers,
thunks, jump-table stubs) rather than payload blocks, that is a claim about what a
CFG-based detector keys on, and it agrees in direction with an independent measurement
on a different corpus.

CFGNet-v2 is the choice when accuracy on novel samples is the criterion and its cost is
acceptable, but choosing it deepens the explainability tension: the most accurate
architecture in the study is also the least locally explainable one. That tension is a
finding, not an oversight, and we leave it open.

\subsubsection{Implications for how detectors are evaluated}

The narrow reading of this paper is a recommendation about operators; the broader one
concerns evaluation practice and is the more consequential.

If an in-distribution benchmark can recommend \emph{against} message passing where a
temporally split benchmark recommends \emph{for} it, such benchmarks are not merely
optimistic by a constant: they can be directionally wrong about which family of models
to build. Discounting reported accuracy by an allowance for temporal bias does not
repair that, because the bias is architecture-dependent. The remedy exists: TESSERACT
\cite{pendlebury2019tesseract} specifies what a temporally sound evaluation requires and
Arp et al.\ \cite{arp2022dosdonts} document how often it is skipped. What we add is a
demonstration, inside CFG-based detection, that skipping it changes the conclusion
rather than only the numbers.

Three secondary implications follow. The temporal split carries design information, not
merely a corrective (Section~\ref{sec:d-rq3}). Explanation studies need more than one
architecture: a study reporting that its explainer selects meaningful subgraphs has, on
this evidence, characterised its operator rather than its data. And both numbers must be
reported: an in-distribution figure alone cannot be converted into a deployment
expectation, and $\Delta$AUC alone can be improved by making the in-distribution number
worse (Section~\ref{sec:r-pretrain}).

\subsection{Comparison with prior work}
\label{sec:d-prior}

Our results agree with the prior literature in places, extend it in others and
contradict it in one.

\paragraph{Where this work agrees} The general finding of TESSERACT
\cite{pendlebury2019tesseract} and Arp et al.\ \cite{arp2022dosdonts}, that
time-agnostic evaluation inflates reported performance, reproduces here in a domain
neither studied: every model in Table~\ref{tab:r-runA} scores lower across the boundary
than within it, by 0.031 to 0.136 AUC. The choice of GraphSAGE by MalGraph
\cite{ling2022malgraph} is independently vindicated: on our corpus GraphSAGE has the
lowest degradation of anything we tested and the lowest training cost, which is a
stronger justification for that design decision than the near-tied accuracies their own
backbone ablation could offer. And the concern that motivates MalGNE's vocabulary-free
encoding \cite{peng2024malgne}, out-of-vocabulary behaviour in dictionary-based
instruction encodings, is the same concern that fixes the node representation used here.

\paragraph{Where this work extends prior results} TESSERACT establishes the protocol
for feature-vector Android detectors and applies it without comparing architectures;
we apply it to twelve graph architectures and find that the inflation is
architecture-dependent, which turns a discount factor into a design consideration.
MalGraph applies a time-based split to malware only; moving the boundary in both
classes is what exposes the rank reversal of Section~\ref{sec:r-flat}, since the
flat control's collapse is measured against benign programs that moved too.
Shokouhinejad et
al.\ \cite{shokouhinejad2025consistency} measure explanation consistency on a single
corpus under synthetic perturbation; we measure it across a real temporal boundary,
add behavioural grounding against a pre-registered effect-size threshold, and add a
cross-architecture agreement test their single-corpus design cannot express. The
survey of Shokouhinejad et al.\ \cite{shokouhinejad2025survey} names temporal
robustness as essentially unevaluated in this literature; this paper is a direct
response to that observation.

\paragraph{Where this work disagrees} The headline claim of the attention-guided
stacking ensemble \cite{shokouhinejad2025stacked} (that stacking three base learners
through an attention meta-learner improves on the base learners and on a plain average)
does not reproduce under a temporal split on our corpus. The ensemble improves on
neither, and loses to a single well-chosen operator (Section~\ref{sec:r-repro}). Two
qualifications belong with that statement. Our reproduction substitutes our
37-dimensional node vector for their autoencoder-reduced encoding, and the base-learner
accuracies land two to four points below theirs in consequence, so the disagreement is
about the \emph{stacking gain} rather than about their absolute numbers. And their paper
does not claim temporal robustness (it names the absence of such an evaluation as a
limitation), so this is a result their design does not contradict so much as leave open.
What our reproduction shows is that the gain is specific to in-distribution evaluation,
and that the meta-learner's weights are nearly input-invariant, which explains why.

\paragraph{The one prior operator comparison across a boundary, and why it concludes
the opposite} MalGraph \cite{ling2022malgraph} ablates its backbone over GraphSAGE,
GCN and TransConv under its time-based split and reports that the architecture ``is not
sensitive to the choice of GNN variants'', at 99.49, 99.93 and 99.97\% AUC. That
contradicts RQ1. Three features of their design account for the difference, and the
protocol used here avoids each. Their split
is time-based for malware only, goodware split randomly at matching proportions, so the
benign half of the shift is removed by construction, and Section~\ref{sec:r-drift}
measures the benign class as drifting at least as much, making the removed half the
larger. Their three variants span 0.48 AUC points above 99\%, a range inside which no
effect of the size we measure could be resolved. And they report no in-distribution
figure, so nothing is differenced: the ablation ranks operators on absolute accuracy
after the boundary, not on how much each lost crossing it. We read their result as
evidence that operator choice does not matter at 183,000 samples, saturated accuracy
and a malware-only boundary, and ours as evidence that it does at 682 graphs,
unsaturated accuracy and a boundary moving both classes. The two findings are therefore
compatible, each within its own regime.

A second, smaller disagreement is internal to the graph-learning literature: GIN's
provable 1-WL expressiveness \cite{xu2019gin} does not translate into temporal
durability here. GIN finishes sixth of nine on 2026 and degrades more ($\Delta=0.073$)
than GCN ($\Delta=0.054$), a strictly less expressive operator. Expressiveness in the
limit and generalisation from a few hundred graphs are different properties, and we
measure the second.

\subsection{Threats to validity}
\label{sec:d-threats}

\paragraph{Confounding of the measured shift} This is the most important limitation. The
benign control drifts at least as much as malware (domain-classifier AUC 0.952 against
0.886), so a toolchain, compiler or collection artefact explains it at least as well as
malware evolution. This does not weaken the comparative claims (all architectures face
the same shift), but the paper says nothing about how detectors respond to malware
evolution specifically.

\paragraph{Scale, corpus provenance and survivor bias} 682 usable graphs from 1,989
binaries is small, and our corpus is purpose-built for the reasons set out in
Section~\ref{sec:why-custom}. The consequence is that our absolute numbers are not
comparable with figures published on EMBER, BODMAS or SOREL-20M, which is why every
claim we make is internal to the corpus and why we reimplemented the closest
published system on it. The 60-second extraction timeout is additionally a sampling
decision, so the corpus over-represents binaries that angr analyses quickly and
under-represents heavily packed and obfuscated programs, the same criticism that
applies to other static pipelines, including MalGraph \cite{ling2022malgraph}.
Per-class yield is comparable (34\% and 30\%), so the loss introduces no large class
bias.

\paragraph{Split variance} Every $\pm$ is over five seeds on one fixed split, and the
paired bootstrap of Section~\ref{sec:stats} covers sampling of the evaluation graphs.
Neither covers the choice of the 293/74/92 partition itself: we did not run repeated
splits, so sensitivity to that partition is unmeasured. This is a documented gap rather
than an oversight, and it bounds how finely the rankings should be read.

\paragraph{Depth and width matching} Depth and width are matched only within the
three-layer group. DeeperGCN runs at seven layers because depth is intrinsic to that
architecture, and the higher-capacity variants carry four to twelve times GCN's
parameters. Parameter count does not predict the ranking (the two largest models finish
seventh and ninth), but the comparison is not parameter-matched and we do not claim it
is.

\paragraph{Power of the isolated readout test} On the GIN backbone the readout swap
gains $+0.084$ AUC at $p=0.095$, against a comparator whose own standard deviation is
0.087: a failure to demonstrate, not a demonstration of absence.

\paragraph{Re-training of the Run B references} Run B re-trained its reference
architectures rather than reloading them, carrying up to 0.006 AUC of GPU
non-determinism, below the seed standard deviation of roughly 0.019 and changing no
ranking, but the cross-run rows of Table~\ref{tab:r-field} carry it in addition to seed
variance.

\paragraph{Explainability coverage} Coverage is partial: sixty graphs, one checkpoint
and one seed per architecture; we did not measure attribution variance across seeds, the
stability result is powered for large shifts only, and GNNExplainer's masking baseline
could select different nodes under a different choice.

\paragraph{Remaining limitations} No family-level analysis is possible, since our
corpora carry no family labels, so every result is binary detection and the
covariate-shift versus concept-drift decomposition cannot be separated empirically. Two
artefacts are descriptive only: the month-wise breakdown of the 2026 corpus, four of
whose seven months are benign-only, and the apparent large-graph improvement in the
size-tercile analysis. Finally, we do not evaluate adversarial robustness: MalAOI
\cite{peng2025malaoi} shows CFG-based GNN detectors of exactly this kind can be evaded
by maliciousness-preserving opcode insertion, and nothing we measured speaks to it.

\section{Conclusion}
\label{sec:conclusion}

We asked whether the way CFG-based malware detectors are evaluated changes which
detector a study recommends, and found that it does.

We trained twelve graph neural network variants and a flat-feature control on Windows
control flow graphs collected in 2024--2025 and scored them once on graphs collected in
2026, under a protocol in which every design decision reads only data from before the
boundary. The message-passing operator changes robustness to the shift significantly, in
two independent executions and across seven pairwise comparisons that survive
multiplicity correction, and every surviving comparison separates an architecture that
builds its graph vector by parametric aggregation from one built around a learned
attentional readout. The principal measurement is a rank reversal: a flat model given
every node feature but no topology is the best in-distribution model in the study and
among the worst across the boundary, so a conventional benchmark would have rejected
message passing outright. Recalibration does not repair the degradation, ensembling
three fragile operators does not substitute for choosing one robust operator, and for
the highest-capacity models more training data makes out-of-distribution performance
worse while making the in-distribution number better.

Attributions survive the same boundary, but explanation validity is a property of the
architecture rather than of the data. Nothing shifts significantly in what the models
attend to, yet the two architectures we studied ground their explanations in
incompatible ways and agree with each other at chance level, and the operator that
scores highest on the later corpus is the one hardest to explain locally. A
single-architecture attribution result in this domain therefore characterises the
operator, not the task.

The drift finding carries design information. CFGNet-v2, which we specified from the
finding rather than found by search, matches the best operator in a twelve-variant field
and beats five comparators under correction, including the flat control. It is a tie
with PNA rather than a win, and its within-run advantages do not all survive an
independent initialisation, but the two architectures we designed on intuition finished
last and second to last, and the one we designed on evidence finished first.

Four qualifications travel with these results. The measured shift is confounded: benign
software drifts at least as much as malware, so these are results about robustness to a
measured distribution shift, not malware evolution. The corpus is small, at 682 graphs,
purpose-built because no public corpus releases raw PE binaries of both classes with
per-sample dates, and biased toward binaries a static disassembler analyses quickly. The
reported $\pm$ is optimisation variance; we bootstrap the evaluation graphs alongside
it, but split variance remains unmeasured. And we did not test adversarial robustness.

We recommend GraphSAGE in practice: within 0.014 AUC of CFGNet-v2 on the later
corpus, the lowest degradation we measured, an eighth of PNA's peak memory, and
attributions that are at least characterisable. Our methodological recommendation is
broader and more durable. A benchmark that does not separate training and evaluation in
time is not merely optimistic: it can be directionally wrong about which family of
models to build, and the correction is not a discount factor applied to reported
accuracy but a different measurement.

\subsection{Future work}
\label{sec:future}

Five directions follow from those limitations. Deconfounding the shift is the most
valuable: a corpus with class-matched collection windows, or one recording compiler and
packer provenance, would separate malware evolution from toolchain change.
Repeated-split intervals would quantify the split variance we leave unmeasured and say
how finely the rankings can be read. Family-level labelling, through a tool such as
AVclass \cite{sebastian2016avclass} over multi-vendor verdicts, would let the
covariate-shift and concept-drift components be separated rather than only named.
Attribution across seeds and explainers would establish whether the
architecture-specificity we report is a property of the operators or partly of
GNNExplainer, which a comparison against PGExplainer \cite{luo2020pgexplainer} and
gradient-based attributions would settle. Finally, adversarial robustness under the same
temporal protocol is open: whether an operator resisting natural shift also resists
deliberate opcode insertion \cite{peng2025malaoi} is not implied by anything we
measured.

\section*{CRediT authorship contribution statement}
\textbf{Md. Asif Sajeed:} Conceptualization, Methodology, Software, Validation,
Formal analysis, Investigation, Data curation, Visualization, Writing: original
draft. \textbf{Md. Nazrul Islam Mondal:} Conceptualization, Methodology,
Supervision, Project administration.
\textbf{Md Ashraful Hossen Akash:} Investigation, Validation, Writing: review
and editing.

\section*{Declaration of competing interest}
The authors declare that they have no known competing financial interests or
personal relationships that could have appeared to influence the work reported
in this paper.

\section*{Funding}
This research did not receive any specific grant from funding agencies in the
public, commercial, or not-for-profit sectors.

\section*{Data availability}
Malware binaries cannot be redistributed, so the corpora themselves are not
deposited. The extractor, training and evaluation code, together with its
version and the frozen node-feature schema, is available at
\url{https://github.com/Ho9pe/TG-CFG} (commit e2b6a55) \cite{tgcfg2026code}.
The extracted control flow graphs and 37-dimensional node features for both
the 2024--2025 training corpus and the 2026 evaluation corpus, identified by
extractor configuration hash \texttt{6bbe4a8649a76a3f}, are deposited at
Kaggle \cite{tgcfg2026data}.

\section*{Acknowledgments}
The authors thank MalwareBazaar~\cite{malwarebazaar} for providing access to the
malware corpus used in this study.

\section*{Declaration of Generative AI and AI-Assisted Technologies}
During the preparation of this work, the authors used generative AI and AI-assisted
technologies to improve grammar, readability, and formatting. After using these
tools, the authors carefully reviewed and edited the content and take full
responsibility for the contents of the published article.

\bibliographystyle{elsarticle-num}
\bibliography{refs}

\end{document}